\documentclass[twocolumn,twocolappendix]{aastex63}
\pdfoutput=1 %
\usepackage[T1]{fontenc}
\usepackage{ae,aecompl}
\usepackage[utf8]{inputenc}
\usepackage[english]{babel}
\usepackage{amsmath,amstext}
\usepackage{apjfonts}
\usepackage[figure,figure*]{hypcap}
\hypersetup{linkcolor=red,citecolor=blue,filecolor=magenta,urlcolor=magenta}
\usepackage[hang]{footmisc}
\received{November 16, 2021}
\revised{January 17, 2022}
\accepted{January 21, 2022}
\submitjournal{ApJ}

\shorttitle{xSAGA I. Satellite Radial Profiles}
\shortauthors{Wu et al.}
\graphicspath{{./}{figures/}}

\begin{document}

\title{Extending the SAGA Survey (xSAGA) I: \\Satellite Radial Profiles as a Function of Host Galaxy Properties}

\correspondingauthor{John F. Wu}
\email{jowu@stsci.edu}

\author[0000-0002-5077-881X]{John F. Wu}
\affiliation{Space Telescope Science Institute, 3700 San Martin Drive, Baltimore, MD 21218, USA}

\author[0000-0003-4797-7030]{J.~E.~G. Peek}
\affiliation{Space Telescope Science Institute, 3700 San Martin Drive, Baltimore, MD 21218, USA}
\affiliation{Department of Physics \& Astronomy,
Johns Hopkins University, 3400 North Charles Street, Baltimore, MD 21218, USA}

\author[0000-0002-9599-310X]{Erik~J. Tollerud}
\affiliation{Space Telescope Science Institute, 3700 San Martin Drive, Baltimore, MD 21218, USA}

\author[0000-0002-1200-0820]{Yao-Yuan Mao}
\altaffiliation{NASA Einstein Fellow}
\affiliation{Department of Physics and Astronomy, Rutgers, The State University of New Jersey, Piscataway, NJ 08854, USA}

\author[0000-0002-1182-3825]{Ethan~O. Nadler}
\affiliation{Carnegie Observatories, 813 Santa Barbara Street, Pasadena, CA 91101, USA}
\affiliation{Department of Physics \& Astronomy, University of Southern California, Los Angeles, CA 90007, USA}

\author[0000-0002-7007-9725]{Marla Geha}
\affiliation{Department of Astronomy, Yale University, New Haven, CT 06520, USA}

\author[0000-0003-2229-011X]{Risa~H. Wechsler}
\affiliation{Kavli Institute for Particle Astrophysics and Cosmology and Department of Physics, Stanford University, Stanford, CA 94305, USA}
\affiliation{SLAC National Accelerator Laboratory, Menlo Park, CA 94025, USA}

\author[0000-0002-3204-1742]{Nitya Kallivayalil}
\affiliation{Department of Astronomy, University of Virginia, 530 McCormick Road, Charlottesville, VA 22904, USA}

\author[0000-0001-6065-7483]{Benjamin~J. Weiner}
\affiliation{MMT/Steward Observatory, 933 N. Cherry St., University of Arizona, Tucson, AZ 85721, USA}

\keywords{
\href{http://astrothesaurus.org/uat/416}{Dwarf galaxies (416)},
\href{http://astrothesaurus.org/uat/594}{Galaxy evolution (594)}, 
\href{http://astrothesaurus.org/uat/595}{Galaxy formation (595)}, 
\href{http://astrothesaurus.org/uat/597}{Galaxy groups (597)}, 
\href{http://astrothesaurus.org/uat/1938}{Convolutional neural networks (1938)}
}

\begin{abstract}
We present ``Extending the Satellites Around Galactic Analogs Survey'' (xSAGA), a method for identifying low-$z$ galaxies on the basis of optical imaging, and results on the spatial distributions of xSAGA satellites around host galaxies.
Using spectroscopic redshift catalogs from the SAGA Survey as a training data set, we have optimized a convolutional neural network (CNN) to identify $z<0.03$ galaxies from more distant objects using image cutouts from the DESI Legacy Imaging Surveys.
From the sample of $>100,000$ CNN-selected low-$z$ galaxies, we identify $>20,000$ probable satellites located between 36--300~projected kpc from NASA-Sloan Atlas central galaxies in the stellar mass range $9.5<\log(M_\star/M_\odot)<11$.
We characterize the incompleteness and contamination for CNN-selected samples, and apply corrections in order to estimate the true number of satellites as a function of projected radial distance from their hosts.
Satellite richness depends strongly on host stellar mass, such that more massive host galaxies have more satellites, and on host morphology, such that elliptical hosts have more satellites than disky hosts with comparable stellar masses.
We also find a strong inverse correlation between satellite richness and the magnitude gap between a host and its brightest satellite.
The normalized satellite radial distribution between 36--300~kpc does not depend on host stellar mass, morphology, or magnitude gap.
The satellite abundances and radial distributions we measure are in reasonable agreement with predictions from hydrodynamic simulations.
Our results deliver unprecedented statistical power for studying satellite galaxy populations, and highlight the promise of using machine learning for extending galaxy samples of wide-area surveys.
\end{abstract}

\section{Introduction} \label{sec:intro}

Satellite galaxies are important testbeds for galaxy formation models.
Modern numerical simulations predict that dozens of luminous satellites surround a typical massive host galaxy like the Milky Way \citep[MW;][]{Zolotov+12,Schaye+15,Wetzel+16,Hopkins+18,Pillepich+18,Agora}.
However, predictions for their number counts and properties depend on physical processes like stellar winds, supernovae, interactions with their host galaxy properties, and cosmic reionization \citep[e.g.,][]{Brooks&Zolotov14,Garrison-Kimmel+17,Kim+18,Tollerud&Peek18,Graus+19,Webb&Bovy20,Green+21a}, which are modeled with ``subgrid’’ prescriptions in current hydrodynamic simulations \citep{Somerville&Dave15,Bullock&Boylan-Kolchin17}.

Meanwhile, the spatial distribution of satellites around central galaxies provides a complementary test of dark matter and galaxy formation physics \citep[e.g.,][]{Kravtsov+04a,vandenBosch+05,Nagai&Kravtsov05,Zentner+05,Tollerud+08,Anderhalden+13,Spekkens+13,Wechsler&Tinker18,Stafford+20,Pawlowski21}.
In particular, the radial distribution of satellites around hosts encodes information about the halo accretion history, tidal interactions, and gas stripping \citep{Garrison-Kimmel+17,Richings+20,Samuel+20,Carlsten+20}.
However, the Local Group only features two massive galaxies, the MW and M31, which hinders statistical  comparisons  with  theory.
Furthermore, the radial profiles of MW and M31 satellites are known to differ \citep{McConnachieIrwin06,Richardson+11, Yniguez+14}.

Satellite galaxies are often faint and have low surface brightness, complicating observational identification.
In recent years, deep spectroscopic surveys have begun to identify satellites in the halos of nearby MW \textit{analogs}.
These surveys are designed to complement studies of the satellite systems of the MW and M31, which have been characterized in incredible detail down to the ultrafaint dwarf regime \citep[e.g.,][and references within]{McConnachie12,Martin+16,DrlicaWagner+20}. %
The Satellites Around Galaxy Analogs Survey \citep[SAGA;][]{SAGA-p1} is one such campaign that studies satellite populations in a large sample of MW analogs at distances of $D =$ 25--40 Mpc.
While the SAGA program requires tremendous amounts of observing time in order to measure the redshifts of all satellite galaxy candidates, it is able to achieve unprecedented spectroscopic completeness, enabling a detailed study of satellites brighter than $M_r < -12.3$.
SAGA is already revealing discrepancies between the star-forming fraction of relatively bright satellites in the Local Group and those in other nearby halos \citep{Wetzel+15,SAGA-p1,SAGA-p2}.%

Deep imaging surveys can probe large swaths of sky with fantastic sensitivity, enabling new analyses of fainter and lower-surface brightness satellites around Local Volume and low-redshift hosts \citep[e.g.,][]{Greco+18,Muller+19,Carlsten+20,Carlsten+21b,Martinez-Delgado+21,Tanoglidis+21}.
Wide-area surveys also enable the study of galaxies in environments different from the Local Group, such as satellites in low-$z$ groups or clusters, or alternatively low-mass systems in low-density and isolated environments \citep{SDSS,PanSTARRS,DES,HSC,DELVE}. 
Imaging surveys must be paired with spectroscopic coverage from SDSS \citep{SDSS}, GAMA \citep{GAMA}, and other large surveys in order to confirm the redshifts of satellite and host galaxies.\footnote{For very nearby systems ($D \lesssim 10$ Mpc), surface brightness fluctuations or resolved stellar populations can be used to measure galaxy distances \citep[e.g.,][]{Garling+21,Greco+21,Mutlu-Pakdil+21}. In this work, we will investigate more-distant galaxies ($D = 50 - 130$ Mpc).}
While future spectrographs will be able to acquire thousands of spectra simultaneously \citep[e.g.,][]{DESI} for faint ($r \sim 21$) satellites, current studies with SDSS are only able to probe the brightest confirmed dwarf galaxies \citep[e.g., $M_r \sim -20$ satellites around complete samples of MW analogs;][]{Sales+13}.

We present first results on machine learning-identified low-redshift galaxies using the xSAGA (``Extending the SAGA Survey'') method.
A convolutional neural network (CNN) is trained on SAGA redshifts in order to identify whether candidate galaxies are at low redshift on the basis of three-band optical imaging from the Dark Energy Spectroscopic Instrument (DESI) Legacy Imaging Surveys \citep{Dey+19}.
The xSAGA sample provides an opportunity to study the properties of satellite galaxies across a much larger fraction of the sky compared to existing surveys (i.e., $\sim 1000\times$ the area of the fully completed SAGA Survey), resulting in a massive boost in statistical power.
We characterize the machine learning selection function in detail in order to understand biases, contamination, and incompleteness issues.
In this work, we aim to study the satellite radial distributions for $0.01 < z < 0.03$ host galaxies in the mass range $9.5 < \log(M_\star/M_\odot) < 11$ using the xSAGA CNN-selected sample.

The paper proceeds as follows.
In Section~\ref{sec:data}, we describe the archival data sets used in our analysis.
We describe the machine learning algorithm used for identifying xSAGA low-$z$ galaxies and our methodology for analyzing neural network predictions in Section~\ref{sec:methodology}.
In Section~\ref{sec:profiles}, we investigate the radial distributions of satellites and its dependence on host galaxy properties; readers interested in our results may skip directly to this section.
In Section~\ref{sec:sims}, we compare our results against a limited set of hydrodynamic simulations.
In Section~\ref{sec:conclusions}, we discuss our conclusions and some future work.
All reported magnitudes are corrected for Galactic extinction using the \cite{SchlaflyFinkbeiner11} calibration.
We assume a flat $\Lambda$CDM cosmology with $H_0 = 70~{\rm km~s}^{-1}~{\rm Mpc}^{-1}$ and $\Omega_m = 0.3$, such that an angular separation of $1\arcmin$ corresponds to 12.3 and 36.0 projected kpc at $z=0.01$ and $z=0.03$, respectively.
We also refer to $z < 0.03$ as low-$z$ throughout this paper.
{All of the code used in our analysis is publicly available on Github: \url{https://github.com/jwuphysics/xSAGA}; the version of the code used in this paper can be accessed here: \href{https://doi.org/10.5281/zenodo.5749755}{10.5281/zenodo.5749755}.}

\section{Data} \label{sec:data}

We present an overview of the data sets used for the xSAGA analysis.
We train a CNN to identify $z < 0.03$ galaxies using the SAGA redshift catalog \citep{SAGA-p2} and corresponding Legacy Survey image cutouts \citep{Dey+19}.
We apply the SAGA~II photometric selection criteria to SDSS photometric catalogs in order to produce a sample of xSAGA photometric candidates, for which we also obtain Legacy Survey image cutouts.
Using the trained CNN, we can make predictions for how likely a candidate galaxy's redshift is at $z < 0.03$ on the basis of imaging, and output a list of CNN-identified \textit{probable} low-$z$ galaxies.
We refer this latter sample as the xSAGA low-$z$ catalog.
A subset of galaxies from the xSAGA low-$z$ catalog are positioned near spectroscopically confirmed $z < 0.03$ host galaxies (taken from the NASA-Sloan Atlas catalog); we designate these as xSAGA satellites and hosts, respectively (see Section~\ref{sec:assigning-satellites} for details about the host--satellite assignment).
A brief summary of the tabular-format data is reported in Table~\ref{tab:data}.

\begin{deluxetable*}{l c r c l}
    \label{tab:data}
    \tabletypesize{\footnotesize}
    \tablehead{
        \colhead{Data set} &
        \colhead{Source} &
        \colhead{$N$} &
        \colhead{Spectroscopic?} &
        \colhead{Notes}
    }
    \startdata
    SAGA II+ redshift catalog (train) & SAGA & 112,016 & $\checkmark$ & Spectroscopic redshifts of SAGA candidates \\
    \tableline
    $z < 0.03$ hosts & NSA \texttt{1\_0\_1} & 50,854 & $\checkmark$ & Includes \texttt{galSpecExtra} mass estimates\\ 
    xSAGA hosts & NSA \texttt{1\_0\_1} & 7,542 & $\checkmark$ & $0.01 < z<0.03$ centrals with $9.5 < \log(M_\star/M_\odot) < 11$ \\
    \tableline
    xSAGA photometric candidates & SDSS &  4,411,096 & & SAGA II photometric selection \\ %
    xSAGA low-$z$ catalog (predictions) & CNN & 111,343 & & xSAGA photometric candidates with $p_{\rm CNN} > 0.5$ \\
    xSAGA low-$z$ galaxies around hosts & CNN & 11,449 & & CNN-selected candidates around xSAGA hosts \\
    \enddata
    \caption{A summary of tabular data sets. Training data, test data, and NSA hosts are described in Section~\ref{sec:data}. CNN predictions and results from matching satellites to xSAGA hosts are described in Section~\ref{sec:methodology}.}
\end{deluxetable*}

\subsection{Photometric selection of candidates} \label{sec:photometric-selection}

We apply \citet[][hereafter SAGA II]{SAGA-p2} quality cuts and photometric selection in order to define a catalog of xSAGA photometric candidates.
We select objects from SDSS~DR16 that have small uncertainties and reasonable colors, and are not flagged as containing saturated pixels or registering bad counts.
These objects are then subject to magnitude, color, and surface brightness cuts:

\begin{align*}
    r &\leq 21.0, \\
    \mu_{r,\rm eff} + \sigma_\mu - 0.7(r - 14) &> 18.5,  \\
    g-r - \sigma_{gr} + 0.06(r - 14) &< ~0.9,
\end{align*}
where $\mu_{r, \rm eff} \equiv r_0 + 2.5 \log [2 \pi (R_{r, \rm eff} / {\rm arcsec})^2]$ is the $r$-band surface brightness derived from extinction-corrected magnitudes and Petrosian $r$-band half-light radii, $g-r$ is the extinction-corrected color, and $\sigma_\mu$ and $\sigma_{gr}$ are the (uncorrelated) uncertainties on the surface brightness and color, respectively.

Very similar photometric cuts have been used for selecting SAGA~II spectroscopic targets (see Sections~3.1 and 4.1 of \citealt{SAGA-p2} for details).
One important difference is that xSAGA candidates are selected using SDSS photometry, while the SAGA~II candidates for spectroscopy have been selected using a merged catalog of Dark Energy Survey, Legacy Survey, and SDSS photometry (see Sections~3.4 and 3.5 of \citealt{SAGA-p2}). 
While we could plausibly use Legacy Survey DR9 photometry for defining the sample of candidates, the Legacy Survey DR9 catalogs contain many shredded components of nearby galaxies and artifacts around bright stars. 
These ``junk'' sources overwhelm our low-$z$ candidate list, so we use the less-affected SDSS photometric catalogs instead.
Technically, the SDSS and Legacy Survey filters have slightly different throughputs, resulting in $\lesssim 0.1$~mag discrepancies for $g$ and $r$ magnitudes measured for typical sources in our sample, which in turn have a small impact on the selection criteria.
We note that these discrepancies only affect a small percentage of the xSAGA candidates, and have also been well-characterized in this color--magnitude regime by the SAGA Survey since they also originally used SDSS photometric catalogs \citep{SAGA-p1}.

For the selection of photometric candidates, we do not apply the fiber magnitude cut (\texttt{FIBERMAG\_R} $\leq 23$) imposed by \cite{SAGA-p1} because we may wish to keep sources with very low surface brightness. 
Although we extend the magnitude depth compared to SAGA ($r \leq 21.0$ versus $20.75$), we do not expand the color and surface brightness criteria in this work.
Future work will explore the impacts of loosening other photometric constraints on the selection of low-$z$ candidates.
We compile the full list of SAGA II-selected candidates across the SDSS sky, comprising 6,265,374 objects (of which 4.4 million have valid Legacy Survey imaging, as we will discuss later). 

\subsection{Training sample of spectroscopically confirmed galaxies} \label{sec:training-sample}

We use an expanded version of the spectroscopic redshift catalog presented in the SAGA II paper \citep{SAGA-p2}, hereafter referred to as the SAGA II+ redshift catalog, for training and cross-validating our machine learning algorithm.
SAGA II+ includes 112,016 spectroscopic targets (of which 48,017 were first observed by SAGA) around 86 Milky Way-mass primary galaxies at distances of 25--40~Mpc; almost doubling the redshift sample presented in \citet{SAGA-p2}. 
The total surface area covered by the SAGA II+ redshift catalog is $\sim 70$ square degrees.
{The SAGA II+ redshift distribution is very similar to the one shown in Figure~3 of \citet{SAGA-p2}, with 99\% of the objects at $z < 1$, and a peak around $z = 0.08$.}
Among the SAGA II+ objects 692 have redshifts $z < 0.01$ and 1,856 have redshifts between $0.01 < z < 0.03$.
These data will be presented as part of the Stage III SAGA data release in the near future (Y.-Y.~Mao et al, in preparation; M.~Geha et al., in preparation).

\cite{SAGA-p2} find that the photometric cuts described in the previous section yield an extremely complete sample of $z < 0.015$ galaxies, while reducing the surface density of interlopers by a factor of five.
Nevertheless, as described in \cite{SAGA-p2}, SAGA also obtains redshift for targets that are outside of the SAGA II photometric cuts to verify its target selection method and ensure survey completeness for $z \lesssim 0.01$.
For our redshift range ($z < 0.03$), the SAGA II cuts retain a large majority (89\%) of the spectroscopically confirmed low-$z$ galaxies, but exclude very high-surface brightness objects and very red objects.
Due to these selection criteria, we expect that our xSAGA catalogs will be also be incomplete in these regimes.

\subsection{Sample of host galaxies}

The NASA-Sloan Atlas (NSA) is a catalog of 640,000 spectroscopically confirmed $z < 0.15$ galaxies crossmatched between SDSS and various other surveys \citep{NSA}.
The NSA catalog removes artifacts and incorrectly labeled objects and provides deblended photometry.
We use version \texttt{1\_0\_1} of the NSA catalog, which includes model-fit Sersic indices and elliptical Petrosian fits to the axis ratio and position angle.\footnote{The data model for NSA version \texttt{1\_0\_1} can be found at \url{https://data.sdss.org/datamodel/files/ATLAS_DATA/ATLAS_MAJOR_VERSION/nsa.html}.}
We also crossmatch the NSA catalog with MPA-JHU value-added catalogs in order to use their stellar mass estimates \citep{Kauffmann+03}.

We construct a catalog of ``massive'' host galaxies using NSA low-$z$ objects with $\log (M_\star/M_\odot) > 9.5$. 
We also filter $z > 0.03$ objects and massive galaxy self-matches from the SAGA II low-$z$ candidates.
It is worth noting that some very massive hosts are in known galaxy clusters. 
In Appendix~\ref{app:biases}, we find that the low-$z$ purity and completeness are poor for these extreme systems, so we will later exclude $\log(M_\star/M_\odot) > 11$ host galaxies.

\subsection{Optical image cutouts}

\begin{figure*}
    \centering
    \includegraphics[width=0.495\textwidth]{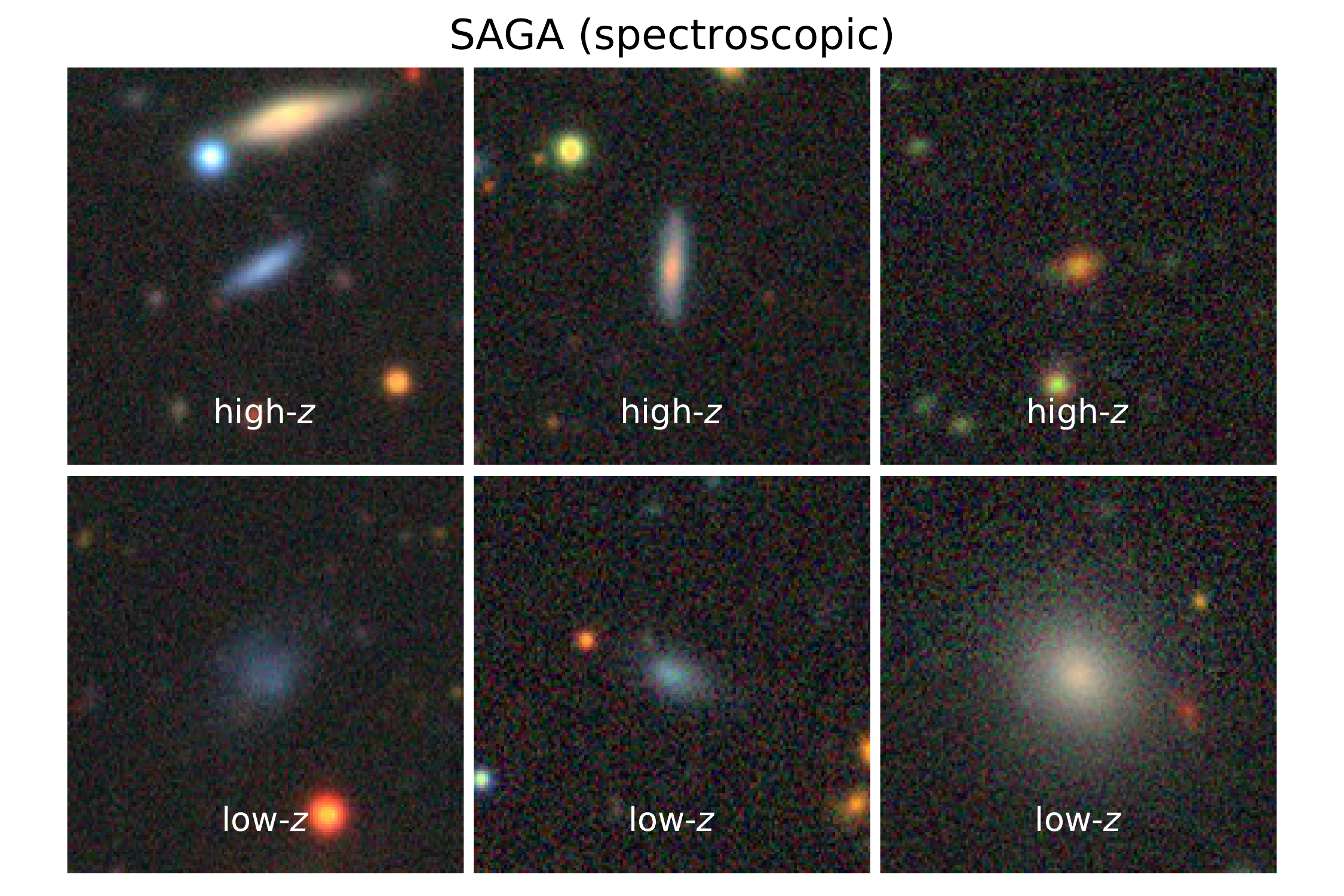}
    \includegraphics[width=0.495\textwidth]{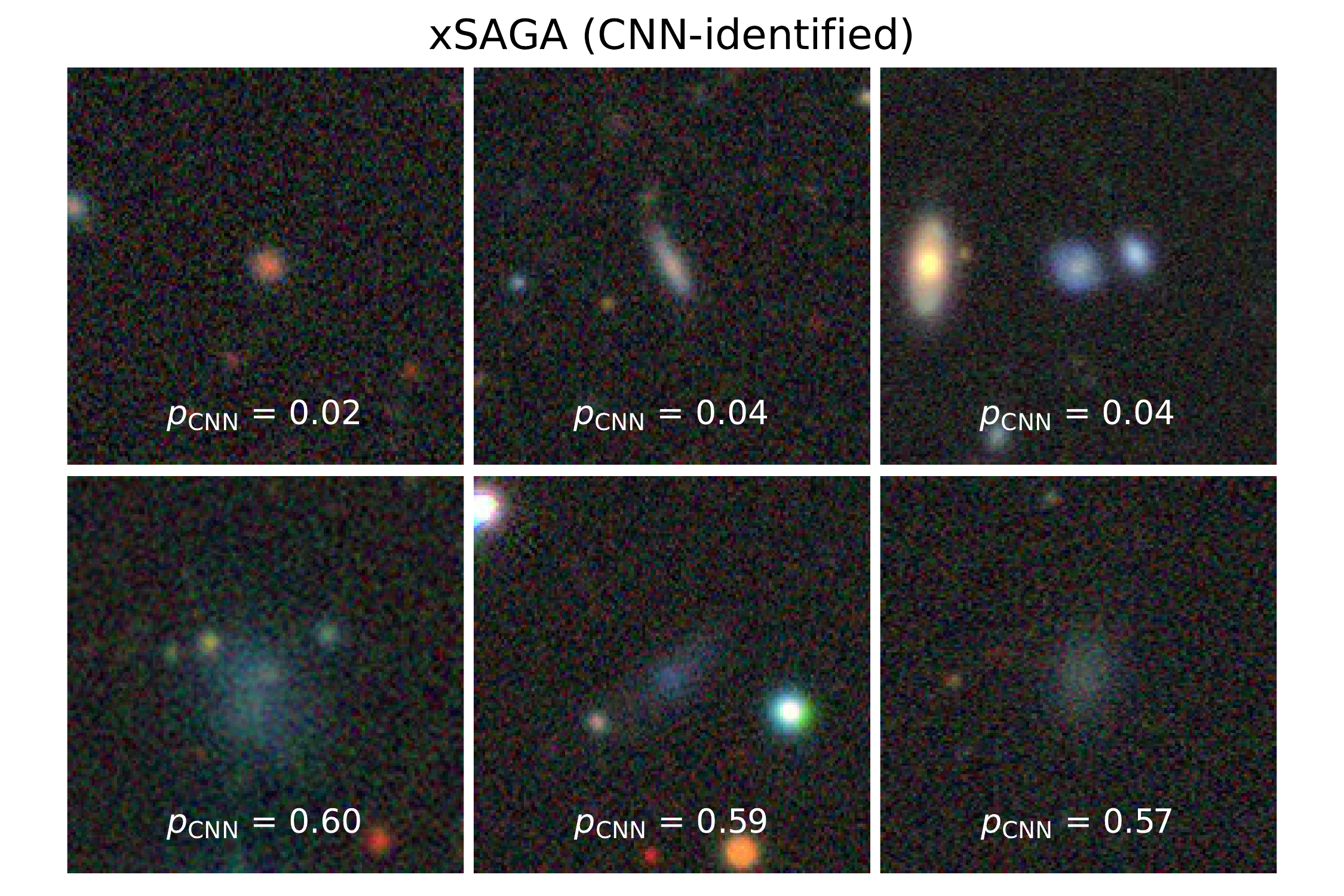}
    \caption{
        Examples of Legacy Survey $grz$ image cutouts corresponding to CNN classifications of {high-$z$ (\textit{upper}) and low-$z$ (\textit{lower})} candidates, for spectroscopically confirmed objects that meet the SAGA-II photometric cuts (\textit{left}; SAGA) and new photometrically selected candidates (\textit{right}; xSAGA).
    }
    \label{fig:examples}
\end{figure*}

The DESI Legacy Imaging Surveys combines imaging from the DECam Legacy Survey, Beijing-Arizona Sky Survey, and Mayall $z$-band Legacy Survey \citep[together referred to as the Legacy Survey;][]{Dey+19}.
Because the Legacy Survey covers most of the contiguous patches of the SDSS, we can obtain deeper imaging for each of the candidates selected from SAGA II cuts in SDSS.

We download image cutouts for the SAGA II+ training data set and for the xSAGA photometric candidates using the Legacy Survey viewer website.\footnote{\url{https://www.legacysurvey.org/viewer}}
We obtain $224\times 224$ pixel $grz$ images at $0.262$ pix~arcsec$^{-1}$ resolution in JPG format, centered on each SDSS object position centroid.
A fraction of our test set is blank, corrupted, or outside the Legacy Survey survey coverage, so we keep only the 4,411,096 valid images with sizes greater than 4096 bytes (out of the original 6.3 million objects in the photometric candidate catalog).
For computational efficiency, all image cutouts are cropped to $144 \times 144$ (i.e., $38^{\prime \prime}$ per side) before being processed by a convolutional neural network.
In Figure~\ref{fig:examples}, we show several example images of SAGA and xSAGA sources.

\section{Methodology} \label{sec:methodology}

We use a CNN, trained on SAGA spectroscopic redshift catalogs, to identify $z < 0.03$ objects on the basis of optical image cutouts (Sections~\ref{sec:cnns} and \ref{sec:crossvalidation}).
We compile a sample of CNN-classified satellites, i.e., probable low-$z$ galaxies around spectroscopically confirmed host galaxies,  check if satellite galaxies selected using our method are biased as a function of host galaxy properties, and remove the systems that exhibit any biases (Section~\ref{sec:assigning-satellites}).
We apply three corrections to our CNN-selected sample, which impact the total abundance and radial distribution of satellites (Section~\ref{sec:corrections}).

\subsection{Convolutional Neural Networks} \label{sec:cnns}

We train a machine learning algorithm to identify low-$z$ galaxies from the large sample of xSAGA photometric candidates.
From the Legacy Survey image cutouts (Figure~\ref{fig:examples}), it is clear that the spectroscopically confirmed low-$z$ and high-$z$ galaxies have distinct morphological features, even though they have been selected using the same photometric cuts.
Therefore, we use a CNN model to classify targets.
Our method is set up as a binary classification problem rather than a regression problem (e.g., estimating an object's distance or redshift) because we are primarily interested in distinguishing foreground galaxies from the far more numerous background galaxies.
We tried to predict the redshifts of all objects in the SAGA~II+ training sample, but found large scatter ($\sigma_{z,\rm pred} \sim 0.05-0.1$), which is unsuitable for our analysis.
Other approaches are better suited for regressing the redshifts of more distant galaxies using photometric information and images \citep[e.g.,][]{Pasquet+19,Hayat+21,Schuldt+21}.

Here, we briefly describe our CNN and optimization routine, and we refer the reader to the detailed explanation in Appendix~\ref{app:cnn-details} for additional information.
Our CNN receives input images and processes them via matrix multiplication and other non-linear transformations.
Each set of transformations is called a ``layer'' in the neural network.
Thus, our CNN can be thought of as a sequence of matched-filter operations (i.e., convolutions) using progressively more complex filters on progressively larger spatial scales, with convolutional filters and other layers comprising the parameters of the model.

Our CNN is based on the \texttt{resnet} architecture \citep{resnets}.
Because there are many more high-$z$ than low-$z$ galaxies in our training data set, we use the Focal Loss function as the optimization objective \citep{FocalLoss}, which is designed for imbalanced classification problems.
As described in Appendix~\ref{app:cnn-details}, we also modify some layers in the CNN in a way that has been shown to improve performance for astronomical deep learning problems \citep[hybrid deconvolution;][]{WuPeek2020}.
The input image data are randomly rotated and flipped in order to help the CNN empirically learn invariances to such transformations \citep[e.g.,][]{Dieleman15}.
Finally, we use an efficient optimizer that allows us to rapidly train a model to convergence \citep{Ranger}.
In summary, our model enhances the standard 34-layer CNN (\texttt{resnet34}) using \textbf{F}ocal \textbf{L}oss, \textbf{h}ybrid \textbf{d}econvolution layers, and other e\textbf{x}tensions to the \textbf{resnet} architecture; we refer to it as the \texttt{FL-hdxresnet34} model.

For each candidate galaxy, the CNN predicts $p_{\rm CNN}$, a measure of how likely the object is to be at $z < 0.03$.
If $p_{\rm CNN} > 0.5$ (a threshold value that we select in Section~\ref{sec:crossvalidation}), then the galaxy is classified as low-$z$; otherwise, it is classified as high-$z$.
During the training phase, we compare the CNN predictions with the spectroscopic labels for galaxies in the SAGA~II+ redshift catalog, i.e., the training data set.
The CNN parameters are updated in a way that minimizes the error according to the implementation described in Appendix~\ref{app:cnn-opt}.
A fraction of the data is withheld for validation; those validation data are not seen during the training process, and are only revealed for understanding the CNN's performance after the training phase is completed.
We find that the CNN is able to perform similarly well on the training and validation data, which indicates that the CNN is able to generalize well, and is not subject to ``overfitting'' on the training data.

\subsection{Cross-validation} \label{sec:crossvalidation}

In order to characterize the trained CNN, we perform cross-validation using the SAGA II+ training data.
We randomly split the SAGA II+ redshift catalog into four equal-sized subsamples, and use 75\% of the data for training the CNN while reserving the final 25\% of the data for validation.
We repeat this four times, permuting the validation subsample each time, so that we can independently evaluate the CNN on all of the SAGA II+ data (this is known as $k$-fold cross-validation, where $k=4$).

After we have selected model hyperparameters and evaluated the cross-validation metrics, we optimize a final \texttt{hdxresnet34} model using the entire training data set. 
The final optimized model is used to predict $p_{\rm CNN}$ for each of the 4.4M images of galaxies in the xSAGA test data set.
Our optimized CNN is also being used to select low-$z$ candidates for spectroscopic follow-up as part of the LOWZ Secondary Target Program in the DESI Survey (E.~Darragh-Ford et al., in preparation).

\begin{figure*}
    \begin{center}
    \includegraphics[width=0.48\textwidth]{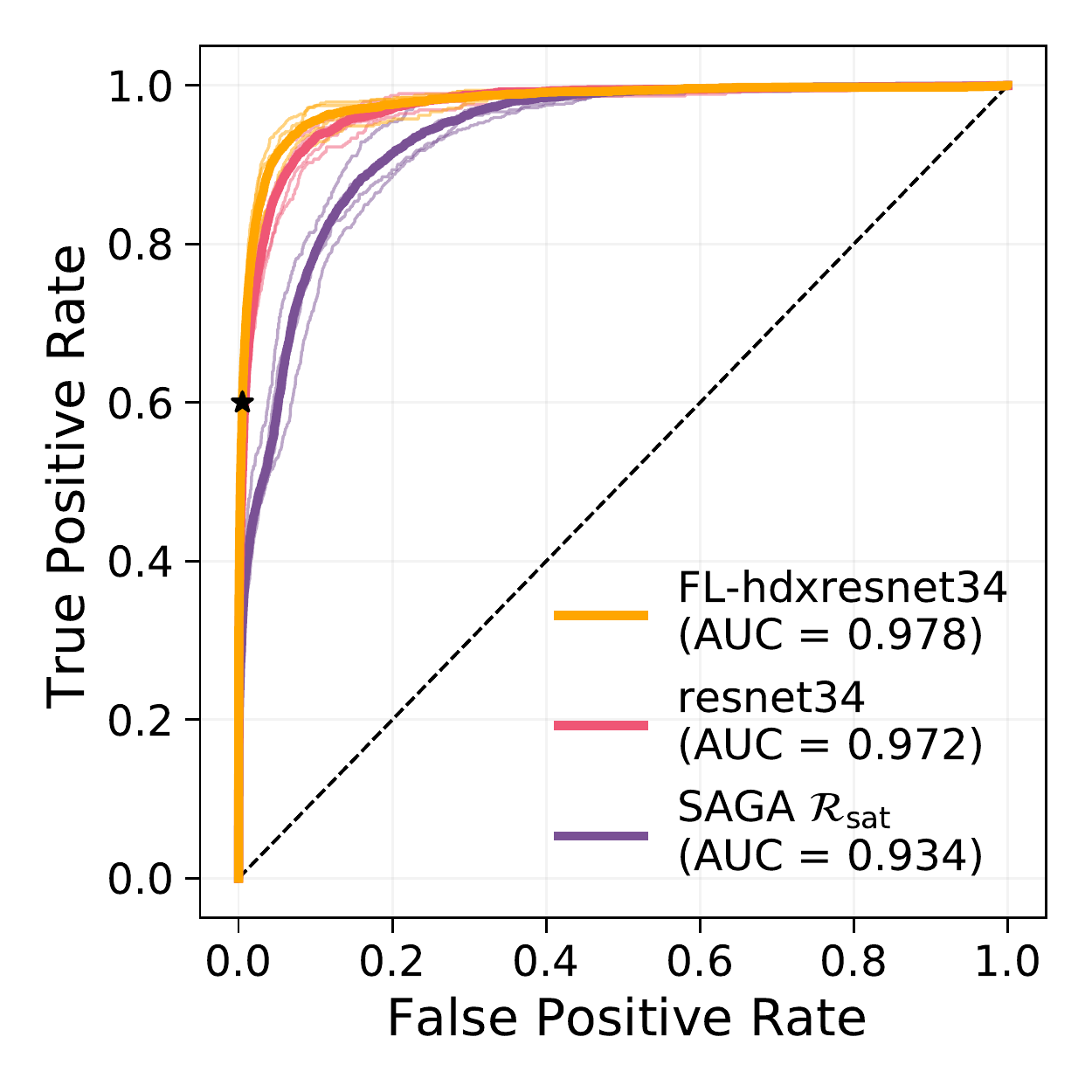}
    \includegraphics[width=0.48\textwidth]{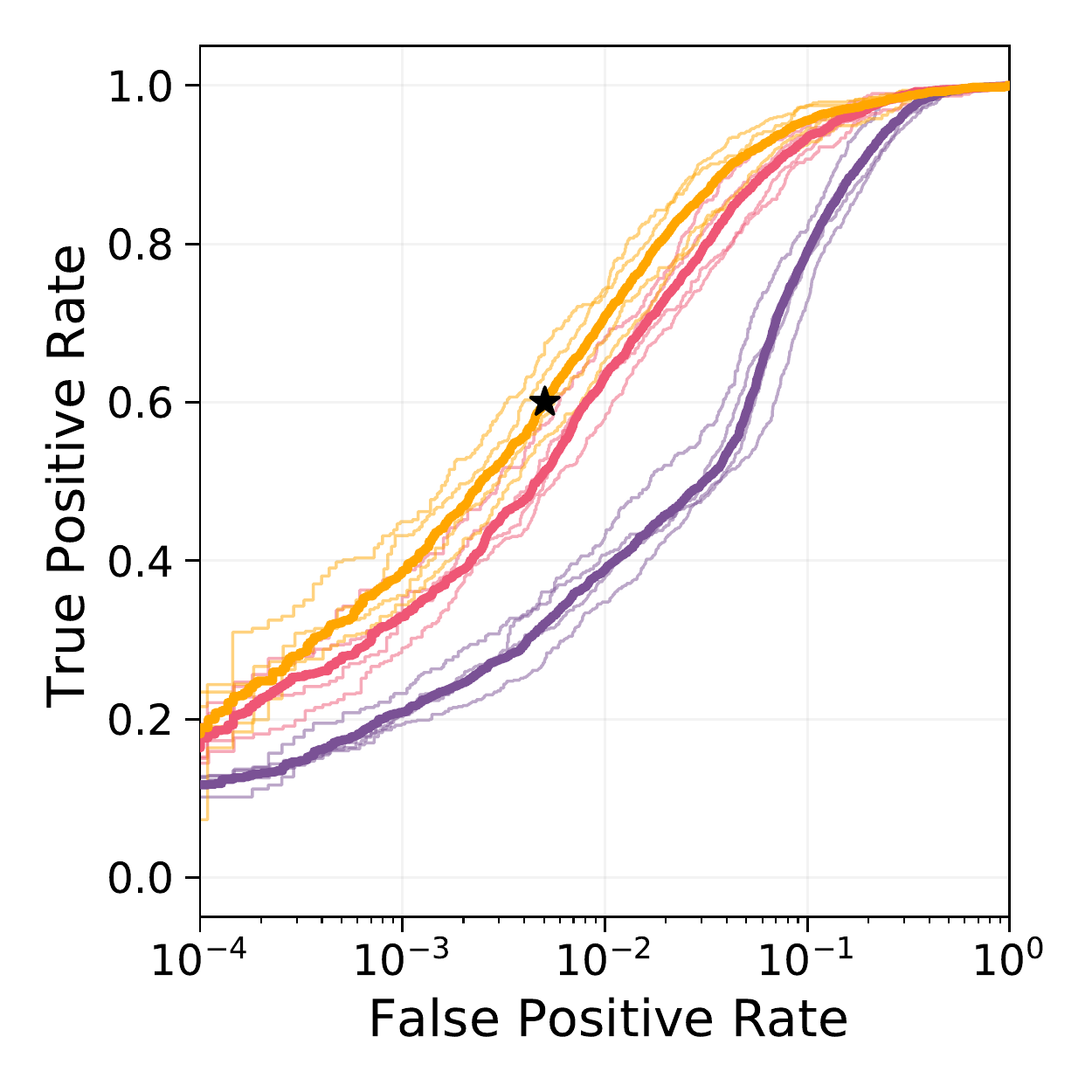}
    \end{center}
    \vspace{-1.5em}
    \caption{
        A {receiving operating characteristic (ROC)} curve for comparing models using the SAGA low-$z$ catalog{, with the false positive rate shown in linear (\textit{left}) and logarithmic (\textit{right}) scale}.
        We show the SAGA~II $\mathcal R_{\rm sat}$ photometric model \citep[purple;][]{SAGA-p2}, a resnet34 CNN model (red), and our enhanced model (yellow), along with their AUC scores.
        The four cross-validation folds are shown as thin lines, while the mean result is shown as a thick line.
        {Our chosen $p_{\rm CNN} = 0.5$ threshold is shown as a black star.}
        A dashed black line shows the expected performance for random guessing (with $\rm AUC = 0.5$).
    }
    \label{fig:cnn-comparison}
\end{figure*}

In order to label low-$z$ and satellite galaxies, we must first determine the threshold probability for classifying a galaxy as a CNN-identified low-$z$ object.
Any choice of $p_{\rm CNN}$ threshold represents a trade-off between the false positive rate (i.e., galaxies that should be high-$z$ but are labeled low-$z$) and the true positive rate.
The purity is defined $\mathcal P \equiv {\rm TP / (TP + FP)}$, and the completeness is defined $\mathcal C \equiv {\rm TP / (TP + FN)}$,\footnote{The purity is also sometimes called precision or positive predictive value and completeness is also called sensitivity, recall, hit rate, or true positive rate.} where TP is the true number of low-$z$ objects, FN is the number of missed low-$z$ objects, and FP is the number of high-$z$ objects incorrectly labeled as low-$z$.

A receiving operator characteristic (ROC) curve displays the false positive rate against the true positive rate for various CNN probability thresholds $p_{\rm CNN}$. 
Models that can achieve low false positive rate without sacrificing the true positive rate will have a large ROC area under the curve (AUC).
In Figure~\ref{fig:cnn-comparison}, we show the ROC curves for a few models evaluated on the SAGA~II cross-validated data.
The AUC is 0.978 for our highly optimized model (\texttt{hdxresnet34}), 0.972 for the base CNN model (\texttt{resnet34}), and 0.934 for the \cite{SAGA-p2} $\mathcal R_{\rm sat}$ model,\footnote{As a caveat, we note that the $\mathcal R_{\rm sat}$ model is only designed to estimate the number of potential SAGA satellites whose redshifts have not been obtained, and should not be considered more than a simple baseline model for comparison to the CNN models. An adapted version of the $\mathcal R_{\rm sat}$ model is discussed in  Appendix~\ref{app:modeling-metrics}.} compared to 0.5 for random guesses, and 1 for a perfect model.
We choose a threshold of $0.5$ for the analysis {(deonted by a black star in Figure~\ref{fig:cnn-comparison}).
Because the data contain many more FPs than TPs, our $p_{\rm CNN}$ threshold does not balance TP and FP rate, and instead results in a similar number of FPs and FNs.}

\begin{figure*}
    \centering
    \includegraphics[width=0.495\textwidth]{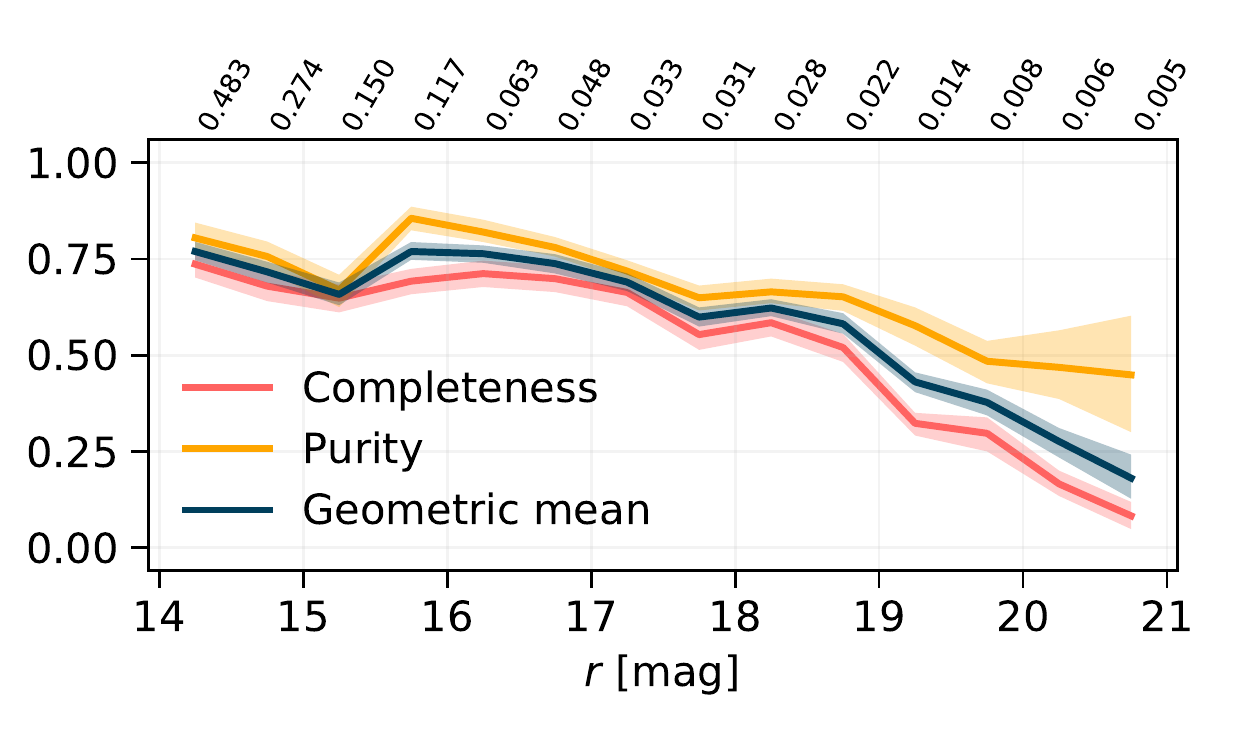}
    \includegraphics[width=0.495\textwidth]{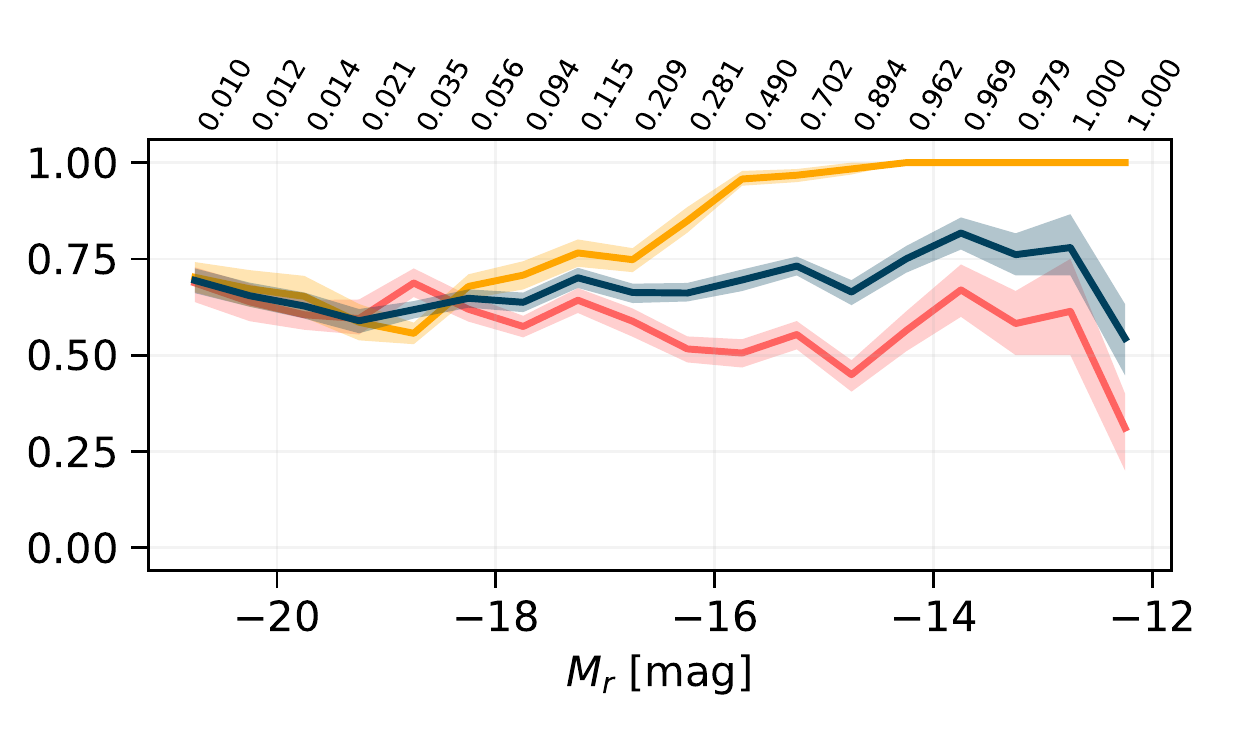}
    \includegraphics[width=0.495\textwidth]{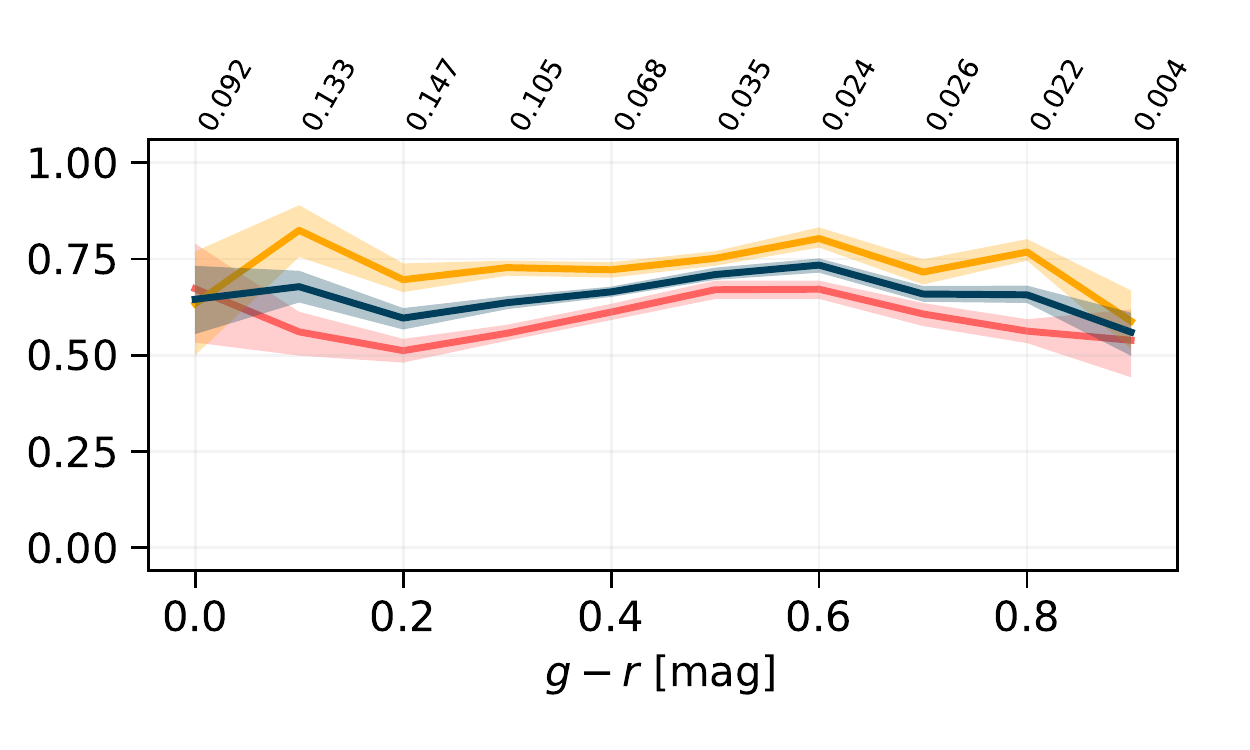}
    \includegraphics[width=0.495\textwidth]{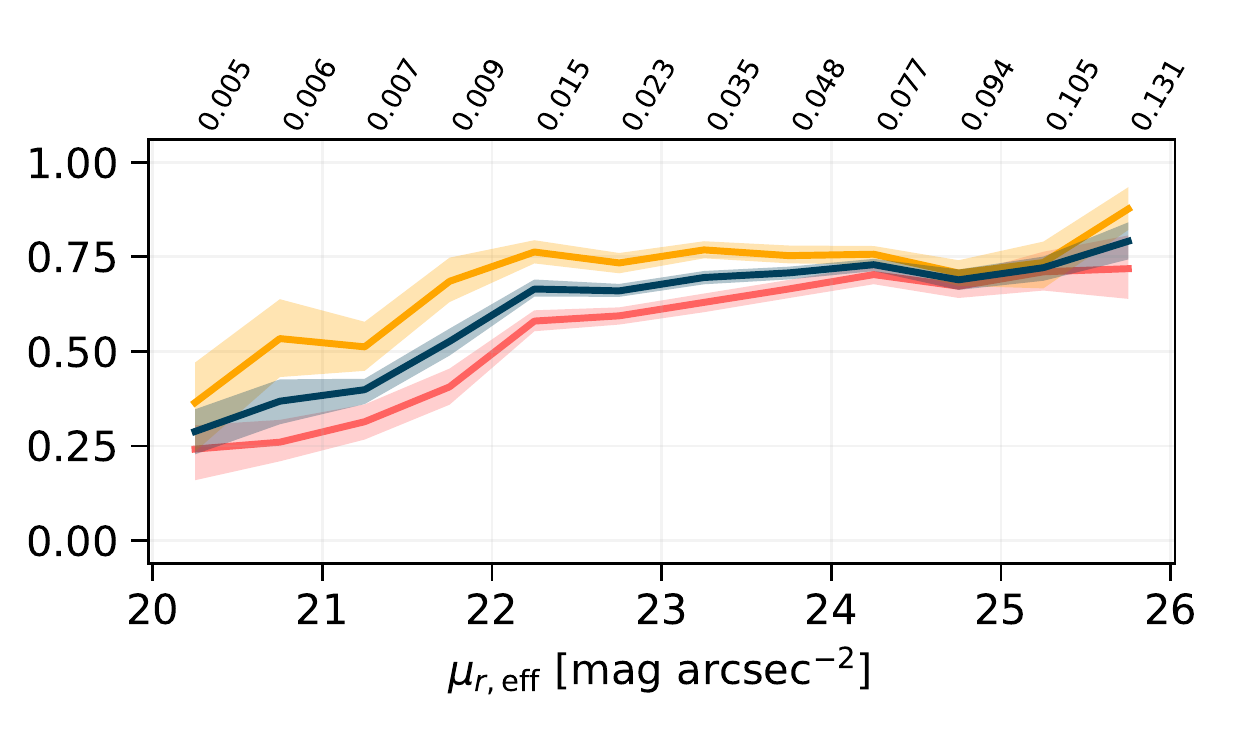}
    \caption{
        Characterizing the performance of the best-fit CNN on the SAGA low-$z$ catalog through cross-validation.
        We example the purity (yellow), completeness (red), and a geometric mean of these two metrics (blue) as a function of apparent magnitude (upper left), absolute magnitude (upper right), color (lower left), and surface brightness (lower right).
        Mean trends and 68\% scatter are computed via bootstrapping the cross-validated data.
        The fraction of true low-$z$ galaxies out of all candidates per bin are shown at the top of each panel.
    }
    \label{fig:cnn-metrics}
\end{figure*}

After selecting a $p_{\rm CNN} > 0.5$ threshold, we evaluate CNN performance using completeness (or recall), purity (or precision), and the geometric mean\footnote{It is also common to consider the harmonic mean (also known as the $F_1$ score) of the completeness and purity, but this essentially tells the same story as the geometric mean.} of the completeness and purity.
For the SAGA cross-validation sample, we report that $\mathcal P = 0.735$ and $\mathcal C = 0.600$.
We compare how these metrics vary with apparent magnitude $r_0$, absolute magnitude $M_{r,0}$, color $(g-r)_0$, and surface brightness $\mu_{r,\rm eff}$, each of which have been corrected for Galactic extinction.
In Figure~\ref{fig:cnn-metrics}, we show CNN metrics as a function of photometric properties.

The cross-validation metrics strongly vary with apparent magnitude and surface brightness.
We find that the CNN exhibits weaker performance in regions of parameter space where truly low-$z$ objects are vastly outnumbered by high-$z$ sources.
For example, only 0.44\% of faint ($r_0 > 20$) sources are at $z < 0.03$, and only 0.65\% of higher-surface brightness ($\mu_{r,\rm eff} < 21 \rm ~mag~arcsec^{-2}$) objects are at low redshift.
In contrast, 6.9\% of $r_0 < 18$ and 5.1\% of $\mu_{r,\rm eff} > 23 \rm ~mag~arcsec^{-2}$ objects are at low redshift.
We also find that the CNN achieves $90\%$ completeness for $z = 0.008$ galaxies (i.e., the redshift of most SAGA sources), compared to $\lesssim 50\%$ for $z > 0.02$ galaxies.
{The completeness generally decreases with spectroscopic redshift out to $z=0.03$.}
These results suggest that the CNN does not perform as well on underrepresented data.
While this kind of bias may be expected, it is important to characterize and take into account.
One solution is to acquire more training data where current data are sparse; in the future, we will be able to alleviate the CNN biases by incorporating data from massively multiplexed spectroscopic surveys such as the DESI Surveys (\citealt{DESI,BGS}; E.~Darragh-Ford et al., in preparation).
In this work, we instead measure the biasing effects of incompleteness and contamination, and empirically devise correction factors for estimating the true number of satellites.

A galaxy at the SAGA spectroscopic completeness limit of $M_r = -12.3$ at $z = 0.01$ has the same apparent magnitude as a $M_r \approx -14.7$ galaxy at $z = 0.03$. 
However, the \cite{SAGA-p2} photometric cuts include $r < 20.75$ candidates, while we have extended the selection to 0.25~magnitudes deeper.
Therefore, we adopt $M_r < -15.0$ as the nominal xSAGA completeness limit.
Figure~\ref{fig:cnn-metrics} shows that the completeness is roughly constant with absolute magnitude, although the cross-validation statistics are noisy.

Finally, we remark that there appears to be another way of measuring CNN performance: by directly interpreting $p_{\rm CNN}$ as a probability. 
If we can assign a probability to each xSAGA low-$z$ prediction, then we may be able to determine correction factors based on these probabilities (e.g., by employing a weighting scheme or by treating CNN identifications as fractional sources).
However, as is well-studied in the literature, deep learning systems are notoriously inaccurate at determining their own uncertainties \citep[see, e.g,][]{UQ}, implying that $p_{\rm CNN}$ cannot robustly be treated as a probability in practice.
Thus, we rely on independent metrics for identifying and correcting different modes of machine learning errors.

\subsection{Assigning CNN-Selected Satellites to Hosts} \label{sec:assigning-satellites}

\begin{figure*}[t]
    \centering
    \includegraphics[width=0.95\textwidth]{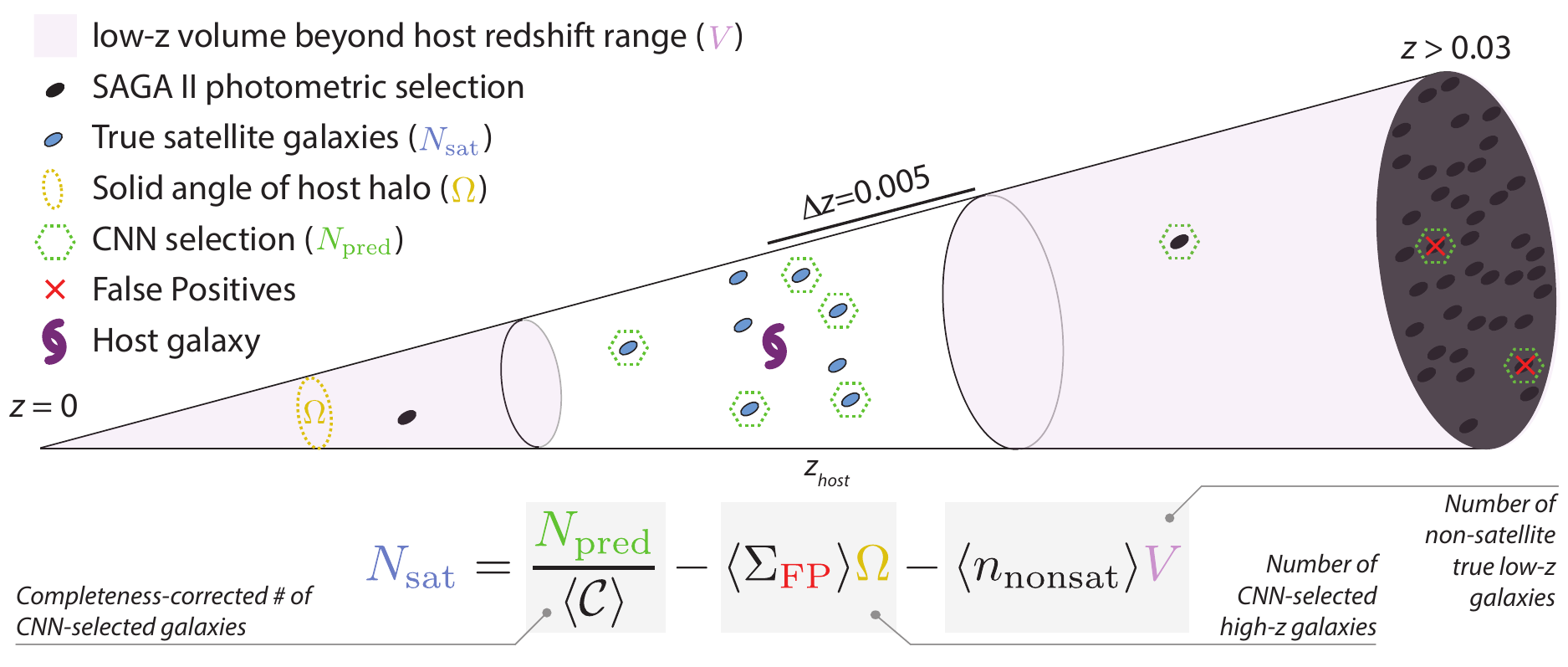}
    \caption{A schematic diagram of our method for estimating the corrected number of CNN-selected satellites (Equation~\ref{eq:corrected-N_sat}).}
    \label{fig:method}
\end{figure*}

We are now able to identify ``probable'' low-$z$ galaxies from the pool of photometric candidates.
This CNN-selected sample includes a variety of objects: isolated field systems, ubiquitous low-mass galaxies in pairs or small groups, and satellites around more massive host galaxies.
We focus our attention on satellite galaxies around spectroscopically confirmed low-$z$ hosts with masses $\log(M_\star/M_\odot) > 9.5$ in this work.

We can determine which CNN-identified galaxies are likely to be satellites around spectroscopically confirmed $z < 0.03$ hosts.
First, we match CNN-selected objects to the closest host galaxies on the sky.
We retain satellites that are between $36$ and $300$~projected kpc from their hosts; the lower bound removes most potentially shredded components of the host galaxy that masquerade as satellite candidates, while the upper bound approximates the virial radius of a SAGA-like host galaxy.
{For the 4,630 CNN-selected galaxies that are assigned to multiple hosts, we only keep the satellite matched to the most massive host.}
We impose a minimum host redshift of $z > 0.01$ in order to ensure that the host halos span less than a degree on the sky.
Very low-redshift hosts are more prone to contamination, as we will discuss later.

Now that we have assigned satellites to hosts, we can test whether the CNN selection is biased as a function of projected separation from the host, host stellar mass, or host morphology.
In order to test the level of bias, we create a simple model for estimating the expected purity and completeness for each CNN-selected satellite on the basis of its photometric properties (see Appendix~\ref{app:modeling-metrics}).
We quantify the CNN bias using the ratio of the modeled purity and modeled completeness, $\mathcal P_{\rm model} / \mathcal C_{\rm model}$, and explore in detail how $\mathcal P_{\rm model} / \mathcal C_{\rm model}$ depends with projected separation from host and with other host properties in Appendix~\ref{app:biases}.
In summary, we do not find significant variation with projected separation, host morphology, or host stellar mass within the range $9.5 < \log(M_\star / M_\odot) < 11$.
For higher-mass hosts, the CNN selection suffers from lower $\mathcal P_{\rm model} / \mathcal C_{\rm model}$, in agreement with previous works that indicated CNN results can be biased for massive galaxies in overdense environments \citep{Wu2020}; therefore we exclude $\log(M_\star/M_\odot) > 11$ hosts from our analysis.
Finally, we ensure that each host is the most massive central galaxy within a projected distance of 1~Mpc and redshift range of $\Delta z = 0.005$ (see Appendix~\ref{app:central-criterion}); these criteria reduce the number of xSAGA hosts from 12,053 to 7,542.
{}

\subsection{Correcting for Incompleteness, False Positives, and Non-Satellites} \label{sec:corrections}

Of the initial 4,411,096 objects, our trained CNN identifies 111,343 probable low-$z$ galaxies above a threshold of $p_{\rm CNN} > 0.5$.
The rate of low-$z$ galaxies in xSAGA (2.52\%) is similar to the rate of low-$z$ galaxies in the SAGA training set (2.28\%). %
Low-$z$ galaxy number counts are corrected using FP statistics and completeness metrics determined from cross-validation tests.
An additional correction term is needed to account for true low-$z$ galaxies that are not satellites of the host (i.e., CNN-selected galaxies within with \textit{projected} separations of 300~kpc, but are in front of or behind the host).
We assume that the FPs and non-satellite galaxies are uniformly distributed with solid angle and proper volume, respectively, and we refer the reader to Appendix~\ref{app:corrections} for more details.

A schematic diagram of our analysis is shown in Figure~\ref{fig:method}.
The fully corrected number of satellites is:

\begin{equation}\label{eq:corrected-N_sat}
    N_{\rm sat} = \frac{N_{\rm pred}}{\langle \mathcal C\rangle} -  \langle \Sigma_{\rm FP}  \rangle \Omega - \langle n_{\rm nonsat}\rangle V ,
\end{equation}
where $\langle \mathcal C \rangle = 0.6$ is the {completeness averaged over the cross-validation sample}, $\langle \Sigma_{\rm FP}\rangle  = 3.04~{\rm deg}^{-2}$ is the surface density of high-$z$ galaxies incorrectly labeled by the CNN as low-$z$ objects, and $\langle n_{\rm nonsat} \rangle = 0.0142~{\rm Mpc}^{-3}$ is the volume density of non-satellite low-$z$ galaxies.
The number of CNN-selected low-$z$ galaxies $N_{\rm pred}$, solid angle $\Omega$, and proper volume $V$ are computed for each individual host in annular radial bins. 
Given these rates, we expect an average (median) of $0.2$ ($0.1$) FPs and $0.4$ ($0.2$) non-satellites per host, with significantly more contaminants for sources at lower redshifts.
In summary, we compute $N_{\rm sat}(<r)$ by summing the cumulative number of CNN-selected galaxies, corrected for incompleteness (see Section~\ref{sec:crossvalidation}), and subtracting the number of expected contaminants, which we assume to be uniformly distributed around the host (see Appendix~\ref{app:corrections}).

{Our CNN selection method has at least an order of magnitude better statistical power than} the traditional approach of counting photometric satellite candidates and subtracting averaged background counts  \citep[e.g.,][]{Holmberg69,Lorrimer+94,Busha+11,Liu+11}.
The CNN selection achieves $60\%$ completeness and $>70\%$ purity, so it is safe to say that the expected signal to noise ratio (S/N) is at least $1$, i.e., we find at least one true low-$z$ satellite for every contaminant.
If we instead only use the photometric cuts described in Section~\ref{sec:photometric-selection} to identify satellite candidates, we find that the S/N $\approx 0.03$.
Alternatively, we can define all $r<21$ objects as satellite candidates.
Although these objects are more uniformly distributed on the sky, they are far more contaminated, and we find that the S/N~$ \approx 0.004$.
Moreover, the S/N will not increase with number of hosts like $N_{\rm host}^{1/2}$, as is expected for shot noise, due to the clustering of background interlopers.
Therefore, when we ``stack'' the hosts together, we expect that the CNN selection should outperform the background subtraction technique in terms of S/N by a factor of \textit{at least} $(1/0.004)^{1/2}$, or $\sim 16$.

\section{Results: Satellite Radial Profiles} \label{sec:profiles}

Using the CNN-identified and corrected sample of satellites, we investigate the radial profile of $M_r < -15$ satellites, i.e., the average number of satellites within our completeness limit as a function of projected host separation.
Satellite radial profiles offer two important observables, both of which may vary with properties of the host galaxies: (\textit{i}) the satellite richness, $N_{\rm sat}(<300~{\rm projected~kpc})$, or the total number of satellites within the virial radius, and (\textit{ii}) the normalized radial distribution of satellites, $N_{\rm sat}( < r)/N_{\rm sat}(<300~{\rm projected~kpc})$, i.e., the shape of the normalized radial profile within 300~projected kpc.
For brevity, we refer to $N_{\rm sat}$ as the total number of satellites within 300~projected kpc.
We analyze 7,542 hosts and 11,449 CNN-selected satellites, which corresponds to a corrected number of 16,241 satellites, for the xSAGA analysis.
We {bootstrap re-sample satellite counts for each radial bin 100 times in order} to estimate uncertainties about mean radial profiles.
Using our large catalog of galaxy hosts, we report how satellite radial profiles depend on host stellar mass, morphology, and brightest satellite properties.

\subsection{Host stellar mass} \label{sec:mass}

\begin{figure*}[t!]
    \centering
    \includegraphics[width=0.495\textwidth]{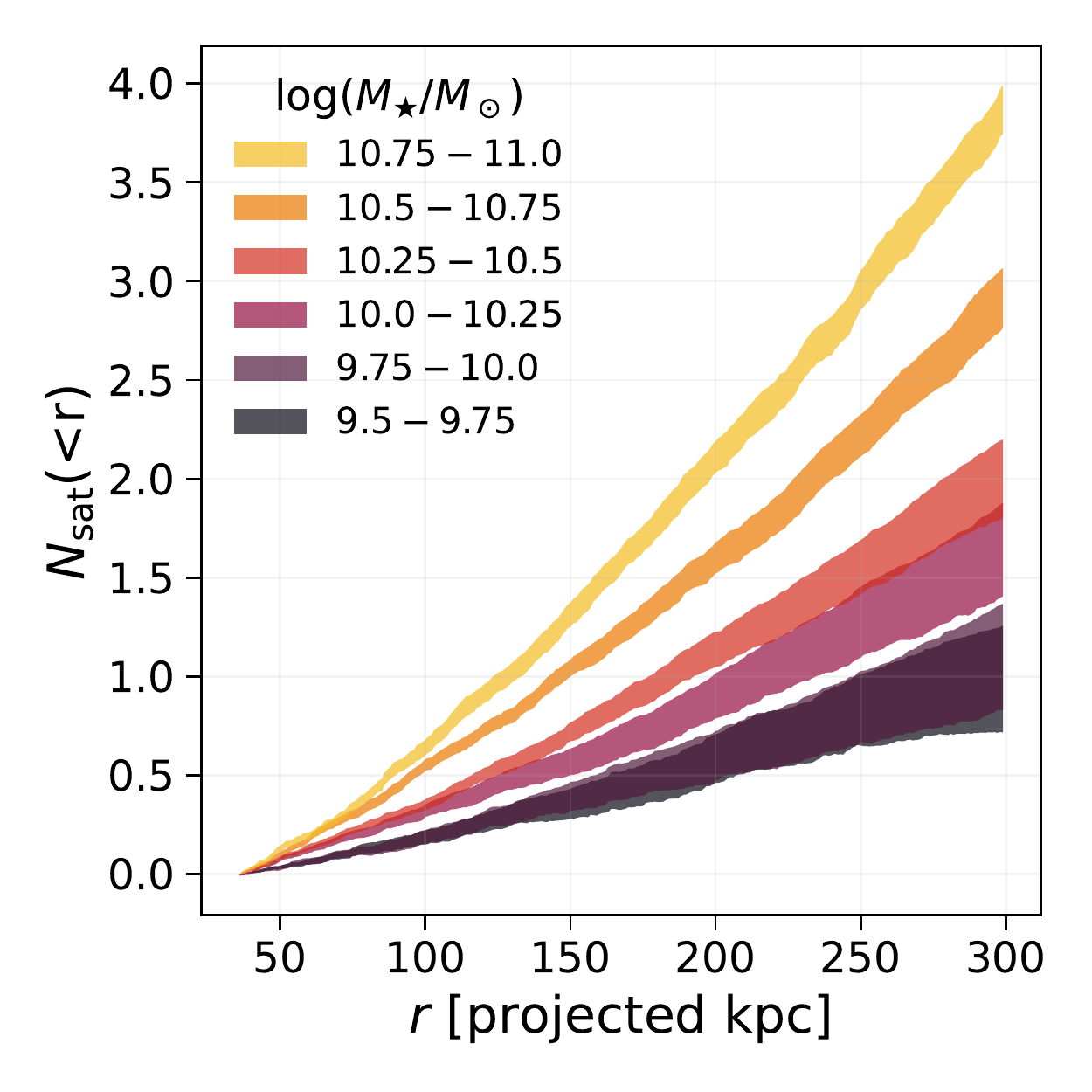}
    \includegraphics[width=0.495\textwidth]{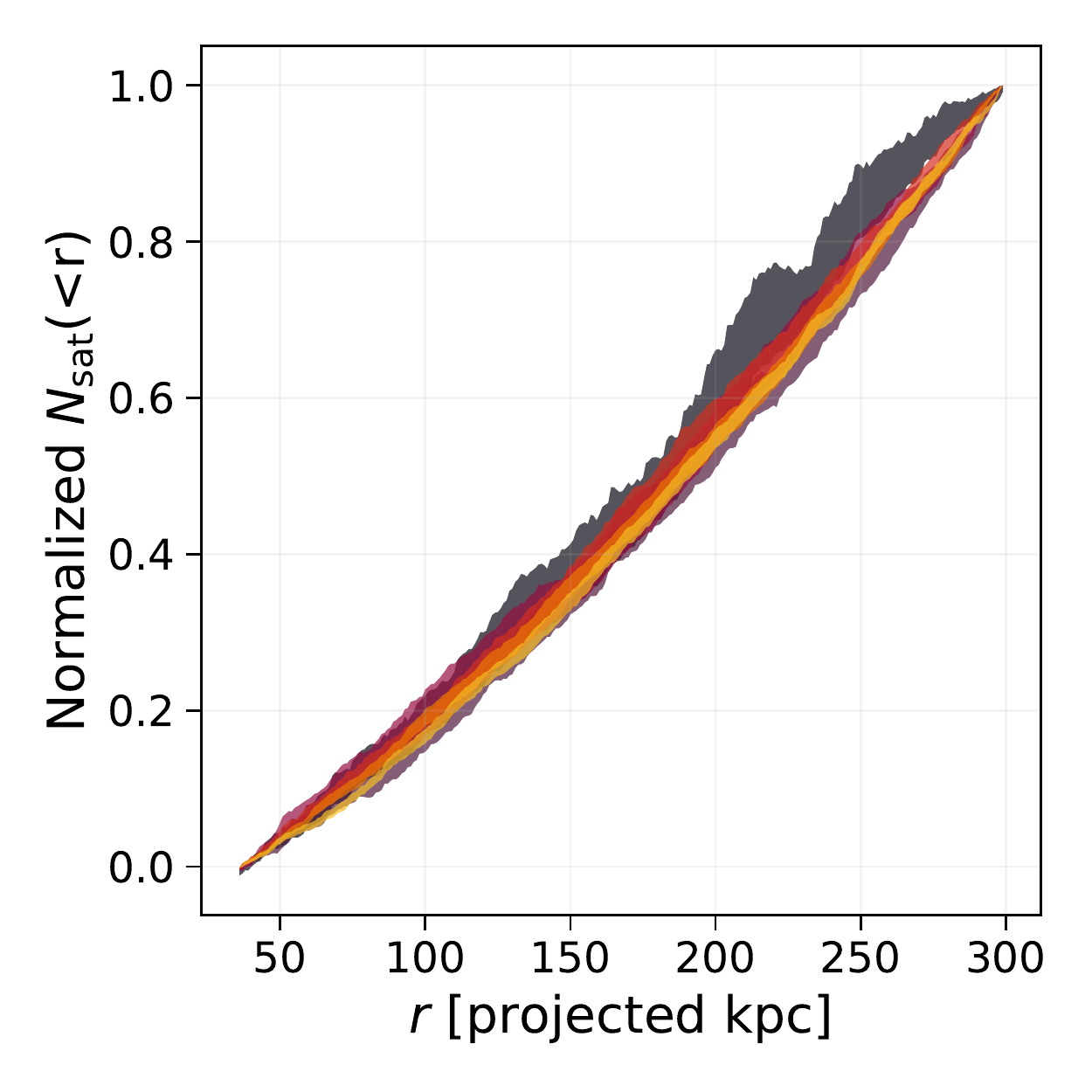}
    \caption{(\textit{Left}) Radial profiles of satellites colored by host galaxy stellar mass in units of $\log (M_\star/M_\odot)$. 
    (\textit{Right}) Satellite radial distributions, i.e., the radial profiles normalized by the total number of satellites within 300~projected kpc, shown for different bins of host stellar mass.
    The intervals depict bootstrapped 16th-84th percentile estimates on the mean radial profiles.
    }
    \label{fig:profile}
\end{figure*}

In Figure~\ref{fig:profile}, we plot the normalized and un-normalized satellite radial profiles in bins of host stellar mass.
The shaded region indicates the bootstrapped standard error on the mean ($\pm 1\sigma$).
Each mass bin contains approximately 900 to 1600 hosts and 900 to 2700 satellites. 
The uncertainties for the mean radial profiles are dictated by the number of samples in addition to the intrinsic host-to-host scatter. 
To highlight this point, we note that there are fewer hosts in the highest mass bin ($10.75 < \log(M_\star/M_\odot) < 11$) than in the lowest mass bin ($9.5 < \log(M_\star/M_\odot) < 9.75$); however, the large variance in satellite profiles for lower-mass hosts is responsible for extra scatter around the mean radial profile.
In Appendix~\ref{app:scatter}, we also explore the impact of binned sample size on the bootstrapped radial profile scatter.

We find that the satellite richness correlates with host stellar mass (see left panel of Figure~\ref{fig:profile}).
Host galaxies in the {Magellanic} Cloud stellar mass range ($10^{9.5} - 10^{10}~M_\odot$) tend to have only one satellite within the xSAGA completeness limits, whereas host galaxies with comparable stellar masses to the MW or M31 ($10^{10.5} - 10^{11}~M_\odot$) have an average of three or four bright satellites. 
Our results agree qualitatively with studies of classical satellite analogs around MW- and M31-like galaxies at higher redshifts \citep{Busha+11,Guo+11,Liu+11,Tollerud+11,StrigariWechsler12,Wang+21}.
Deep observations of nearby faint satellite systems also support this trend \cite[e.g.,][]{Carlsten+20,Carlsten+21a}.
In the right panel of Figure~\ref{fig:profile}, we present the normalized radial distribution of satellites, binned by host stellar mass.
We find that the radial distributions are generally consistent within $\pm 1~\sigma$ scatter.

\begin{figure*}[t]
    \centering
    \includegraphics[width=\textwidth]{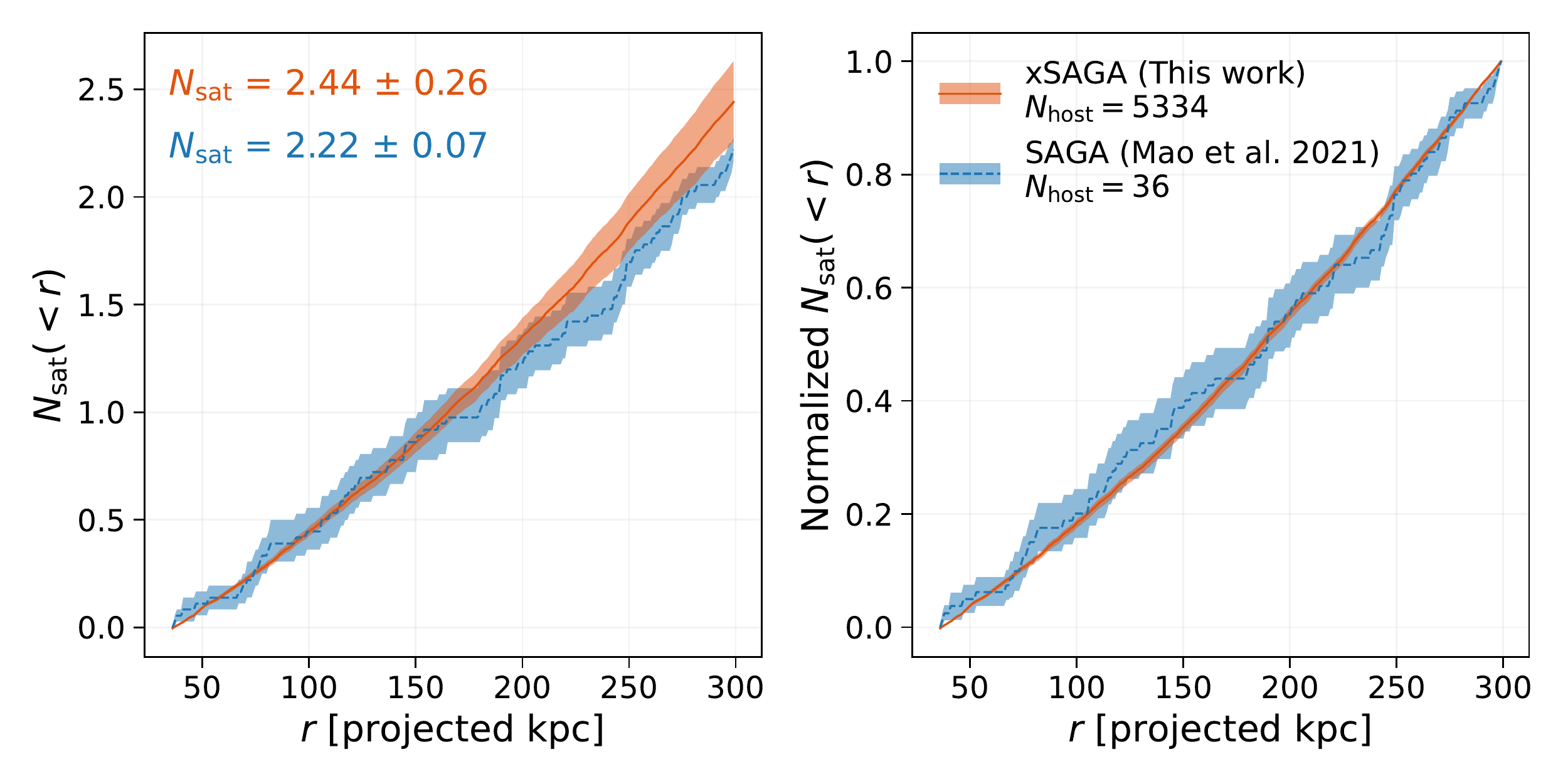}
    \caption{Satellite radial profiles for the xSAGA (orange solid) and SAGA \citep[blue dashed;][]{SAGA-p2} samples.
    We show both the un-normalized radial profiles (\textit{left}) and normalized radial distributions of satellites (\textit{right}).
    The comparison is restricted to satellites within the xSAGA completeness limit ($M_r < -15.0$) and in the SAGA stellar mass range ($10 < \log(M_\star/M_\odot) < 11$).
    We also report the mean number of satellites per host ($N_{\rm sat}$) as well as the number of hosts in each sample ($N_{\rm host}$).
    }
    \label{fig:SAGA-comparison}
\end{figure*}

We also directly compare our findings against the SAGA results \citep{SAGA-p2} using a subset of xSAGA hosts in the same stellar mass range, $10 < \log(M_\star/M_\odot) < 11$.
There are several differences between the SAGA analysis and ours: SAGA satellites (1) have much higher purity, (2) are complete to lower stellar mass, (3) extend to distances below 36~projected kpc.
To ensure a fair comparison, we reiterate that we have corrected for purity, completeness, and backgrounds using the methods outlined in Sections~\ref{sec:methodology}, and only compare against the $M_r < -15$ SAGA satellites in the projected radius range of $36-300$~kpc.

In the left panel of Figure~\ref{fig:SAGA-comparison}, we show the xSAGA and SAGA satellite radial profiles.
We find that the satellite richness for the xSAGA sample slightly exceeds that for SAGA.
However, the difference in $N_{\rm sat}$ is within the bootstrapped error on the mean, while host-to-host scatter is typically larger (for our data, Appendix~\ref{app:scatter}, as well as for the SAGA sample, \citealt{SAGA-p2}, and for simulated samples, \citealt{Samuel+20}).
In the right panel of Figure~\ref{fig:SAGA-comparison}, we show the normalized radial distributions of satellites for xSAGA and SAGA, which appear to be consistent.
We compare the SAGA and xSAGA radial distributions using a two-sample Kolmogorov–Smirnov test in the projected radius range $36$ to $300$~kpc. 
The $p$-value of $0.43$ indicates that the radial distributions are in statistical agreement.

\subsection{Host morphology} \label{sec:morphology}

We investigate the effects of host galaxy morphology on xSAGA satellite radial profiles.
We separate disk galaxy hosts from elliptical hosts on the basis of NSA Sersic index fits to each galaxy's light profile; while not a perfect estimator, Sersic index is sufficient to distinguish disk and elliptical galaxies in broad strokes \citep[e.g.,][]{davari16}. We select $0 < N < 2.5$ for disk galaxies and $3 < N < 6$ for elliptical galaxies.
Overall, more satellites are found around ellipticals than around disk galaxies for the entire xSAGA sample.

\begin{figure}
    \centering
    \includegraphics[width=\columnwidth]{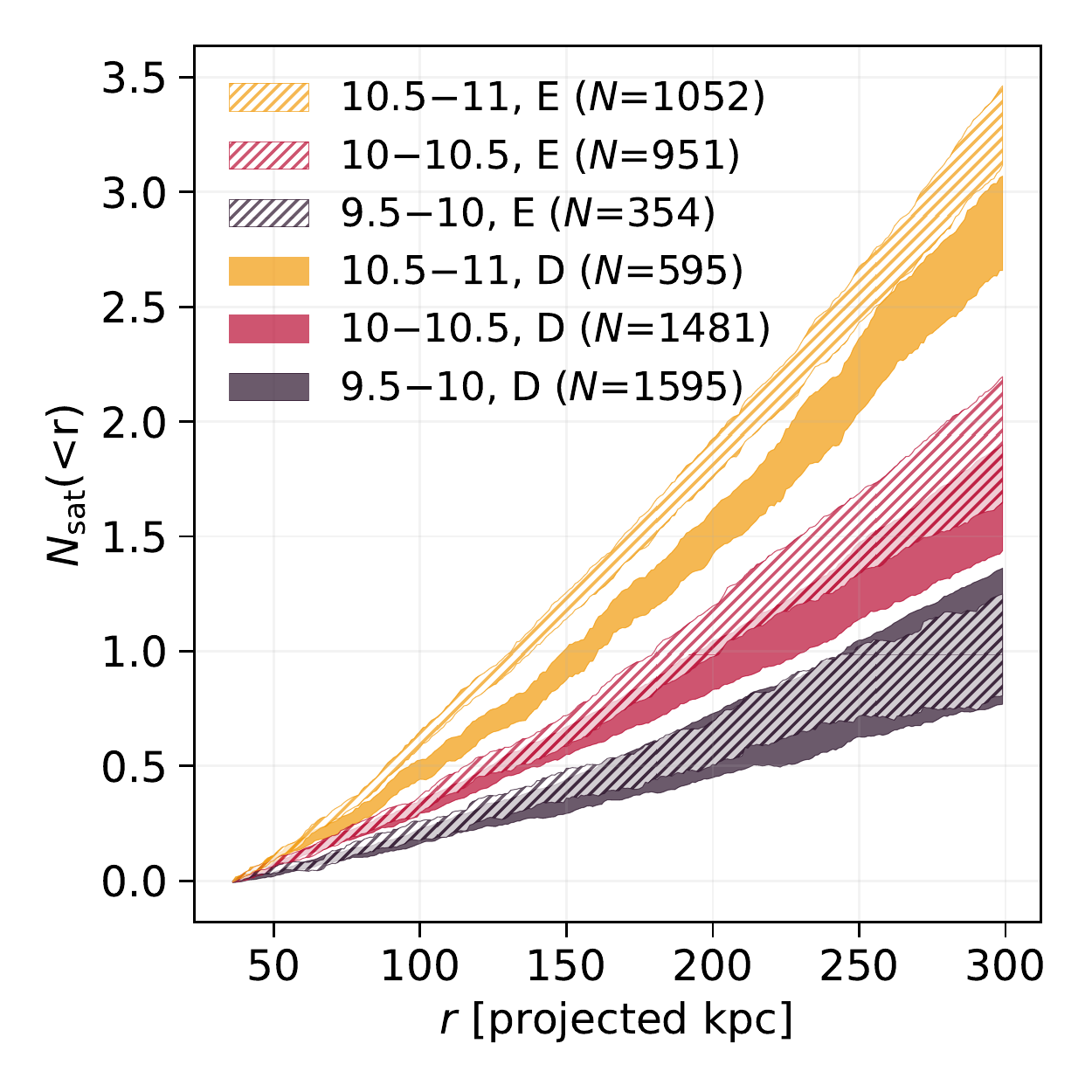}
    \caption{
    {Satellite radial profiles for hosts with disk (D; \textit{solid}) and elliptical (E; \textit{hatched}) morphologies, colored by host galaxy logarithmic stellar mass.
    The legend also describes the number of host galaxies in each bin.}
    }
    \label{fig:morphology}
\end{figure}

Figure~\ref{fig:morphology} shows the satellite radial profiles as a function of host stellar mass and morphology.
In the lowest mass bin ($10^{9.5}-10^{10}~M_\odot$), we find that disk galaxy hosts have approximately the same satellite richness as elliptical galaxy hosts. 
However, higher-mass ellipticals tend to contain more satellites than their disk counterparts.
Our results broadly agree with previous $N_{\rm sat}$ trends with morphology and color \citep[e.g.,][]{WangWhite12,Nierenberg+12,Ruiz+15,Teklu+17}, although some of these studies have reported more extreme differences between the satellite abundances for elliptical and disky hosts outside our mass range ($\log (M_\star/M_\odot) > 11.0$).
It is also worth noting that most SAGA hosts have disky morphologies; if we repeat the comparison between SAGA and xSAGA (left panel of Figure~\ref{fig:SAGA-comparison}) using a host sample that is $80\%$ disks and only 20\% ellipticals, then the xSAGA satellite abundance becomes fully consistent with SAGA ($N_{\rm sat} = 2.21 \pm 0.25$).

\begin{figure*}
    \centering
    \includegraphics[width=\textwidth]{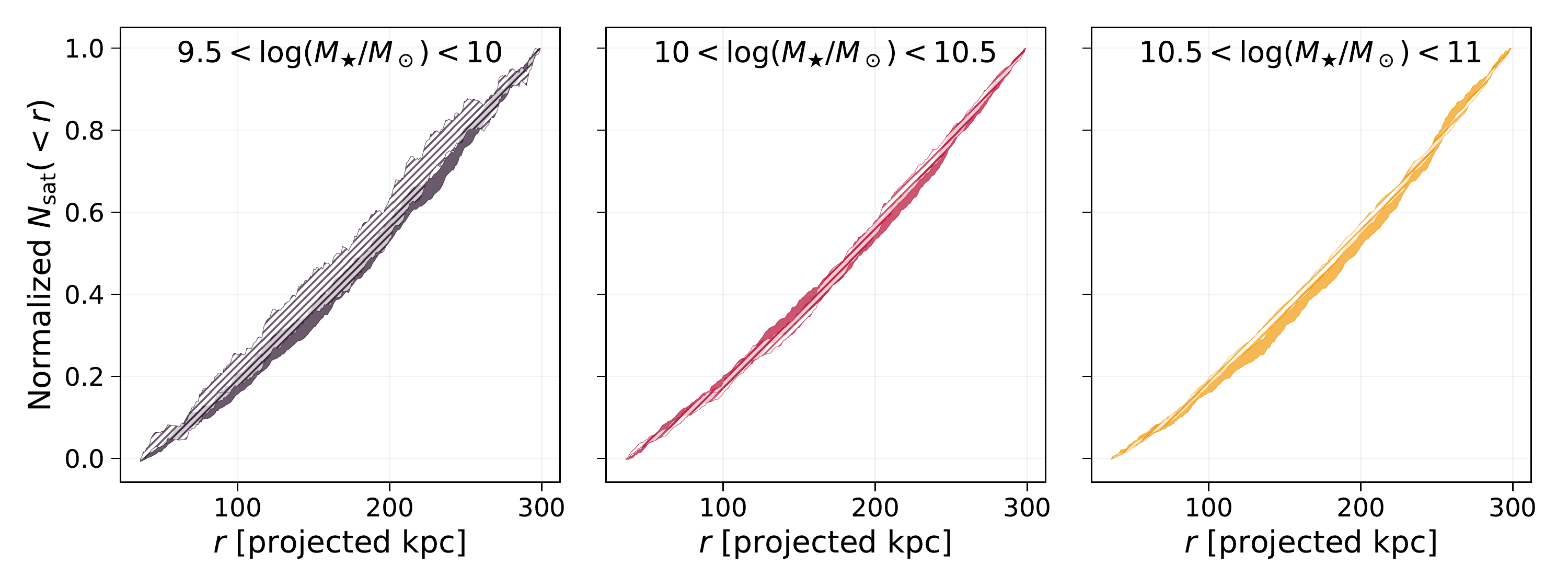}
    \caption{Normalized radial distributions of satellites within 300~projected kpc as a function of host stellar mass (individual panels) and morphology. 
    Similar to in Figure~\ref{fig:morphology}, we divide host galaxies into disky (\textit{solid}) and elliptical (\textit{hatched}) morphologies using their Sersic indices.}
    \label{fig:morphology_radial-dist}
\end{figure*}

In Figure~\ref{fig:morphology_radial-dist}, we show the normalized radial distributions of satellites conditioned on host mass and morphology.
Despite the impact of morphology on satellite richness, we do not find significant differences between the radial distributions for disk and elliptical galaxies.

\begin{figure}
    \centering
    \includegraphics[width=\columnwidth]{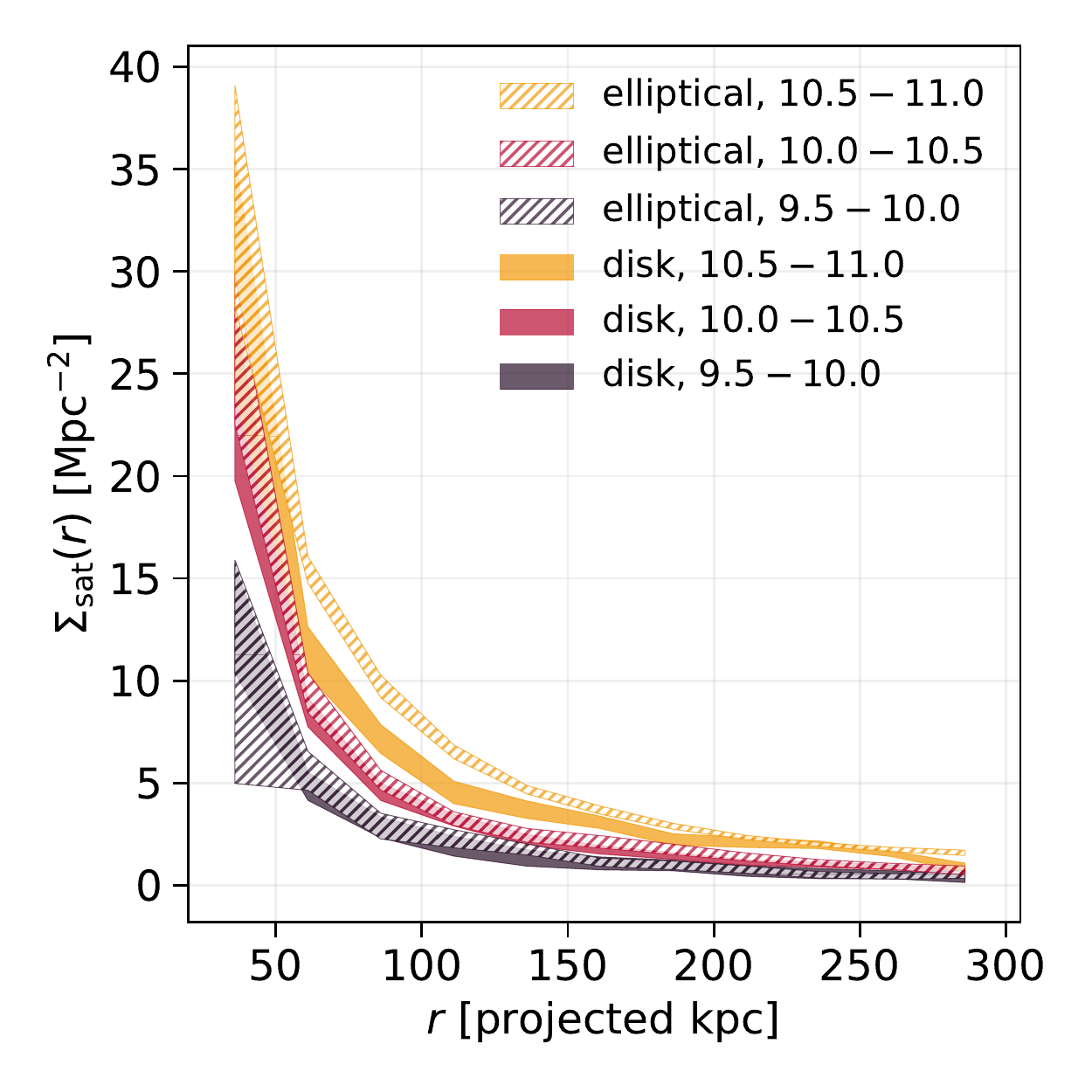}
    \caption{Areal number density of satellites as a function of projected separation from their hosts, presented in the same style as Figure~\ref{fig:morphology}.
    }
    \label{fig:number-density}
\end{figure}

In Figure~\ref{fig:number-density}, we show the projected number density (or areal density) of satellites around hosts of different masses, which is another way of visualizing our results from Figure~\ref{fig:morphology}.
The satellite areal density, $\Sigma_{\rm sat}(r)$, is computed by counting the non-cumulative number of satellites in radial bins, divided by the physical area for each annulus.
As expected from our previous results, the areal density of satellites at any given radius depends on its host galaxy stellar mass, and $\Sigma_{\rm sat}(r)$ is greater overall for elliptical galaxies than for disk galaxies.
However, the satellite number densities do not drop to zero within $300$ projected kpc for hosts of any mass.

Our estimate of the satellite number densities at large radii can be impacted by contamination.
The averaged signal from contaminants should have approximately uniform surface density, according to Equation~\ref{eq:corrected-N_sat}.
We also expect to see non-zero galaxy surface density at $r >300$ kpc because of large-scale structure beyond the host galaxy's halo.
\cite{SAGA-p2} also find several galaxies beyond the host's virial radius (and with separations beyond 300 projected kpc), in addition to some with large peculiar velocities, but label these as field galaxies.
This signal is compounded by projection effects, which impacts more massive and nearby halos. 
Therefore it is unsurprising that our method finds an excess of CNN identifications beyond 300 projected kpc.

It is important to note that 300~kpc is not expected to be the projected virial radius  for many of the xSAGA hosts.
We have chosen this maximum radius in order compare against to the SAGA sample of host galaxies, which {are expected to inhabit dark matter halos $M_{\rm halo} \sim 1.6 \times 10^{12}~M_\odot$}, corresponding to a virial radius $R_{\rm vir} \approx 300$~kpc.
However, the xSAGA host stellar mass range maps to virial masses between $11.5 < \log(M_{\rm halo}/M_\odot) < 12.9$, or $140 < R_{\rm vir} < 400$~kpc.
In Section~\ref{sec:in-virial-radius}, we will present results on satellite populations within the projected virial radii of their hosts.

\subsection{Brightest satellite magnitude} \label{sec:magnitude-gap}

\begin{figure}
    \centering
    \includegraphics[width=\columnwidth]{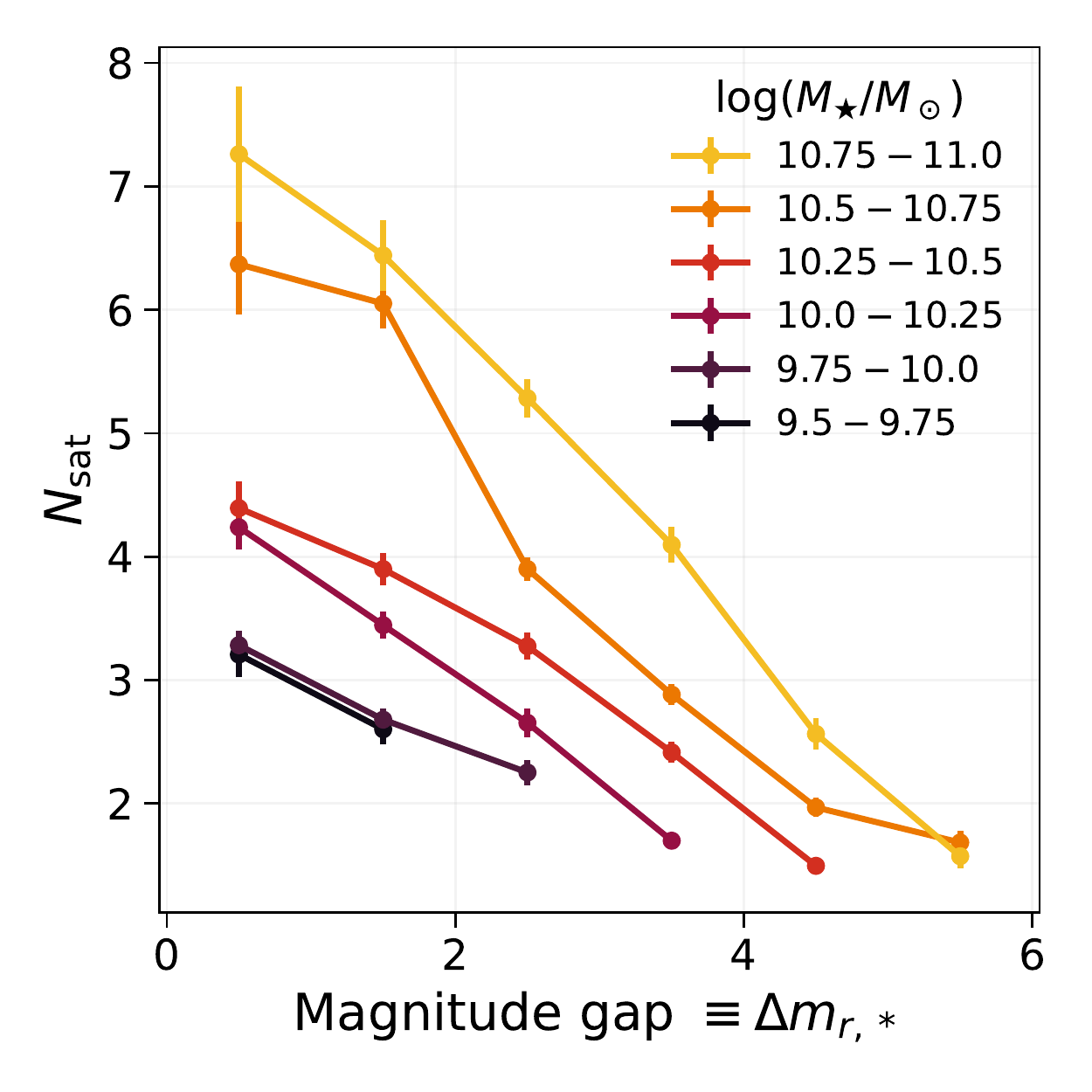}
    \caption{Satellite richness within 300~projected kpc as a function of magnitude gap.
    Host galaxies binned by stellar mass are plotted with different colors.
    }
    \label{fig:satellite-abundance-300kpc}
\end{figure}

The magnitude gap between a host and its brightest satellite galaxy, contains valuable information about the halo accretion history \citep[e.g.,][]{Zentner+05,Mao+15}.
In low-redshift groups and clusters, the magnitude gap between the two brightest members can be used as a secondary halo mass proxy \citep{More12,Hearin+13,Farahi+20}.
Inspired by \cite{Mao+15}, we test how satellite abundances and radial distributions depend on the observed $r$-band magnitude gap, $\Delta m_{r,*}$.

\begin{figure*}
    \centering
    \includegraphics[width=\textwidth]{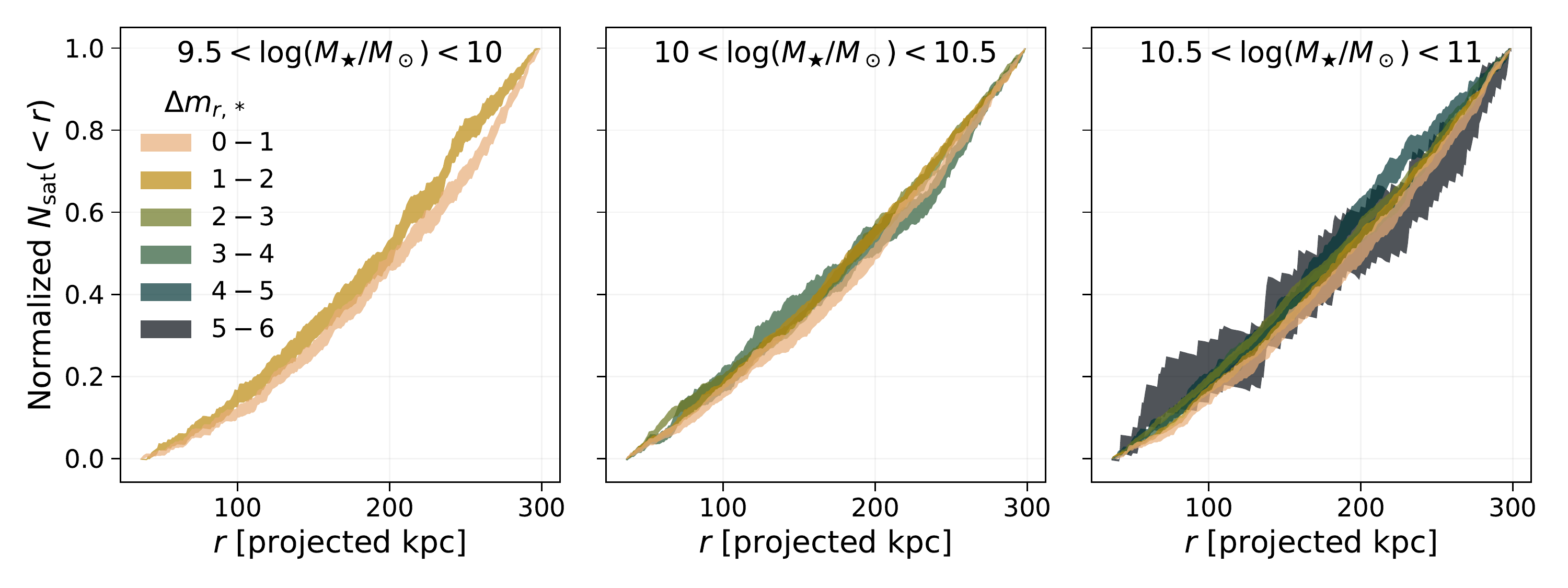}
    \caption{Normalized satellite radial distributions as a function of brightest satellite-host magnitude gap, $\Delta m_{r,*} \equiv M_{r,*} - M_{r,\rm~ host}$.
    }
    \label{fig:radial-distribution_magnitude-gap}
\end{figure*}

We show the dependence of $N_{\rm sat}$ on magnitude gap and host stellar mass in Figure~\ref{fig:satellite-abundance-300kpc}.
While we have already reported the strong dependence of satellite richness with host stellar mass, we now observe a strong anticorrelation between the magnitude gap and satellite richness.
Because we are considering systems that have at least one satellite, the satellite abundances presented here are expected to be higher than for previous results (which are not conditioned on any detected satellites).
A MW-like host ($10.5 < \log(M_\star/M_\odot) < 11$; \citealt{MW-properties}) with an LMC-like galaxy as its brightest satellite ($\Delta m_{r,*} \sim 2$) typically possesses $N_{\rm sat} \sim 6$ detectable satellites within 300~projected kpc.
{If we further restrict to MW-like host galaxies with disky morphologies, then would only expect $N_{\rm sat} \sim 4$ detectable satellites.}
For comparison, the MW has four $M_r < -15.0$ satellites within its virial radius (LMC, SMC, Sagittarius dSph, Fornax), and M31 has six \citep[M33, M32, NGC205, NGC147, NGC185, and IC10;][]{McConnachie12}{; both are in good agreement with our results when we account for xSAGA host-to-host variations}.

We note that the contribution to the satellite richness from additional satellites of the brightest satellite is expected to be small \citep[based on the stellar mass--halo mass relation; see, e.g.,][]{Garrison-Kimmel+17a}; the increase in $N_{\rm sat}$ is more likely driven by host or environmental properties that correlate with the presence of a bright satellite.
Clustered bright contaminants can potentially masquerade as satellites, and may bias our result by increasing both the brightest ``satellite'' magnitude as well as the number of satellites. 
However, we find that hosts with bright CNN-selected satellites have significantly more satellites than similar-mass hosts with comparably bright contaminants, which suggests that the level of bias due to background clustering is small.

We also compare xSAGA radial distributions of satellites as a function of host mass and magnitude gap in Figure~\ref{fig:radial-distribution_magnitude-gap}.
The average distributions do not appear to systematically vary with the magnitude gap, although we observe that the 68\% bootstrapped intervals are noisy and do not fully overlap each other.
We find statistical agreement between all radial distributions conditioned on magnitude gap via two-sided Kolmogorov–Smirnov tests.

\subsection{Satellite abundance within the virial radius} \label{sec:in-virial-radius}

\begin{figure}
    \centering
    \includegraphics[width=\columnwidth]{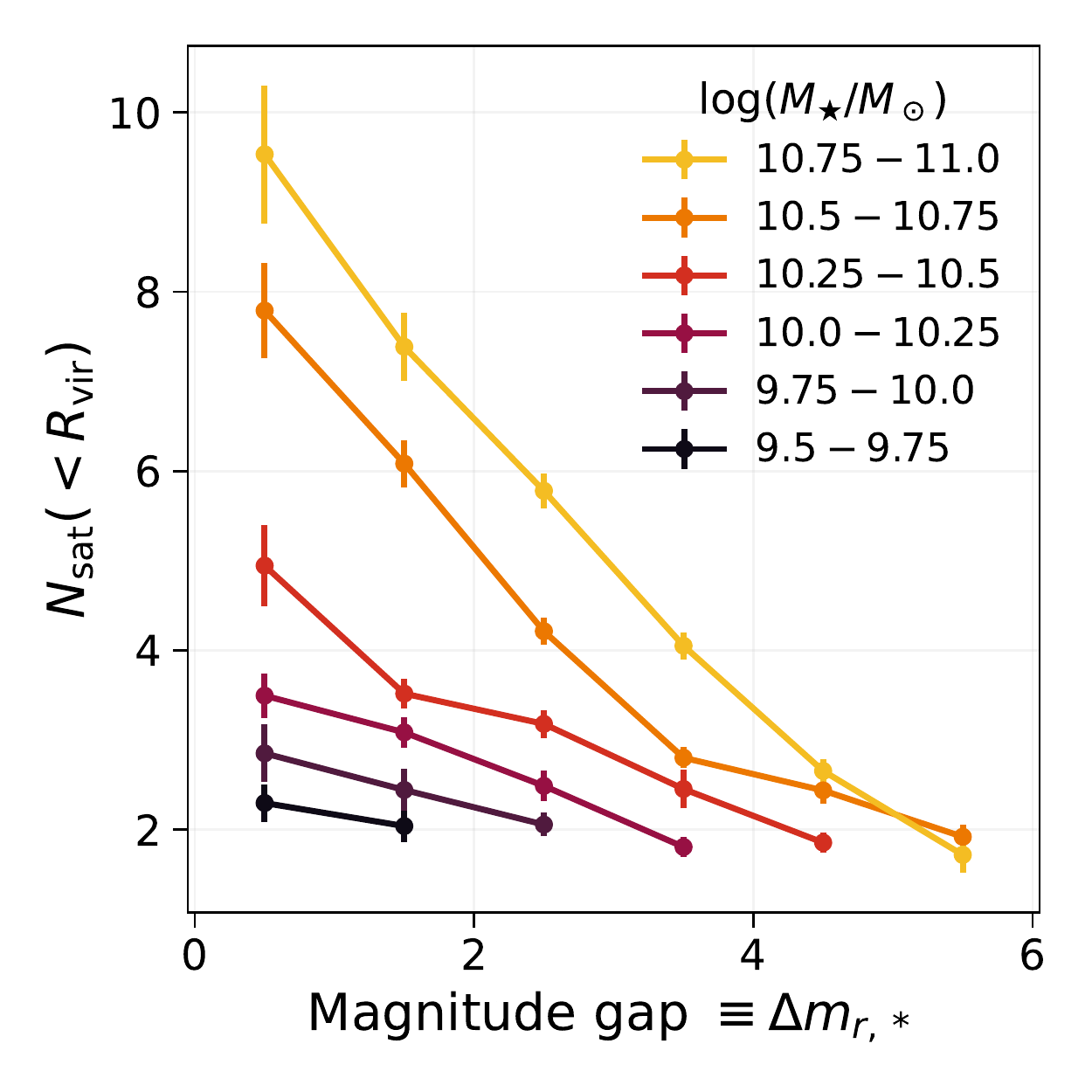}
    \caption{The satellite richness within the virial radius is shown as a function of magnitude gap, conditioned on host galaxy stellar mass.
    }
    \label{fig:satellite-richness_satellite-gap}
\end{figure}

For the radial profiles presented in this work, we have searched for satellites around a constant projected radius of 300~kpc.
This constant radius allows for straightforward comparison against the SAGA Survey, other observations, and numerical simulations.
However, the central and satellite galaxies are theorized to reside in a dark matter halo that is more appropriately characterized by a virial radius or halo mass.
Thus, we explore how the satellite richness within the \textit{virial radius} varies with the stellar mass of the host and the magnitude gap.
We estimate the halo mass using abundance matching results \cite[i.e., a stellar mass--halo mass relation fit to central galaxies;][]{Behroozi+19}, and then derive the virial radius assuming a critical overdensity of $\Delta_{\rm vir} = 99.2$ \citep{BryanNorman98}.
For the host galaxy stellar masses studied here, the virial radius ranges from 140~kpc to 400~kpc.\footnote{Some have argued that satellites may trace the dark matter distribution around central galaxies in a region that extends out to the ``splashback'' radius \citep[e.g.,][]{More+15,Mansfield&Kravtsov20,Diemer21}, defined as the apocenter of dark matter particles in their first orbit around the central halo in a spherically symmetric system. We do not use the splashback radius to define the halo boundary in this work.}
Because the virial radius for $M_\star = 10^{11}~M_\odot$ galaxies exceeds 300~kpc, we redo the assignment of CNN-selected satellites to host galaxies using a maximum separation of 500~kpc, and remove satellite candidates with projected separations greater than their hosts' virial radii.

Figure~\ref{fig:satellite-richness_satellite-gap} shows $N_{\rm sat}(<R_{\rm vir})$ against the magnitude gap, colored by the host galaxy stellar mass.
This trend is stronger than the relationship between $N_{\rm sat}(<300~\rm projected~kpc)$ and magnitude gap (Figure~\ref{fig:satellite-abundance-300kpc}), evidenced by the greater dispersion in satellite abundances.
Our results are unsurprising, since satellite populations should extend out to the virial radius rather than a constant 300~kpc.
We observe that $N_{\rm sat}(<R_{\rm vir})$ increases more quickly with decreasing $\Delta m_{r,*}$ for more massive ($M_\star > 10^{10.5}~M_\sun$) host galaxies.

Until now, we have required that satellites have projected separations of greater than 36~kpc in order to prevent contamination from shredded components of the host galaxy.
However, for galaxies in our stellar mass range, 36~kpc is equivalent to 26--9\% of the virial radius; thus, we have excluded a non-constant fraction of each host galaxy's halo.
To check whether this inner radius cut biases our results, we re-compute the satellite abundance only using satellites within $0.26 < r / R_{\rm vir} < 1.0$.
We find that the satellite abundance systematically decreases for host galaxies of all stellar masses, although there is a larger reduction of satellites around the more massive galaxies with smaller $\Delta m_{r,*}$.
For more massive host galaxies, the satellite abundance within $(0.26-1) \times R_{\rm vir}$ has a shallower dependence on magnitude gap compared to our previous result shown in Figure~\ref{fig:satellite-richness_satellite-gap}.
The normalized radial distributions of satellites within $0.26 < r / R_{\rm vir} < 1.0$ also do not vary with host stellar mass, which is consistent with our findings in Figure~\ref{fig:profile} (right panel).

\begin{figure*}
    \centering
    \includegraphics[width=0.49\textwidth]{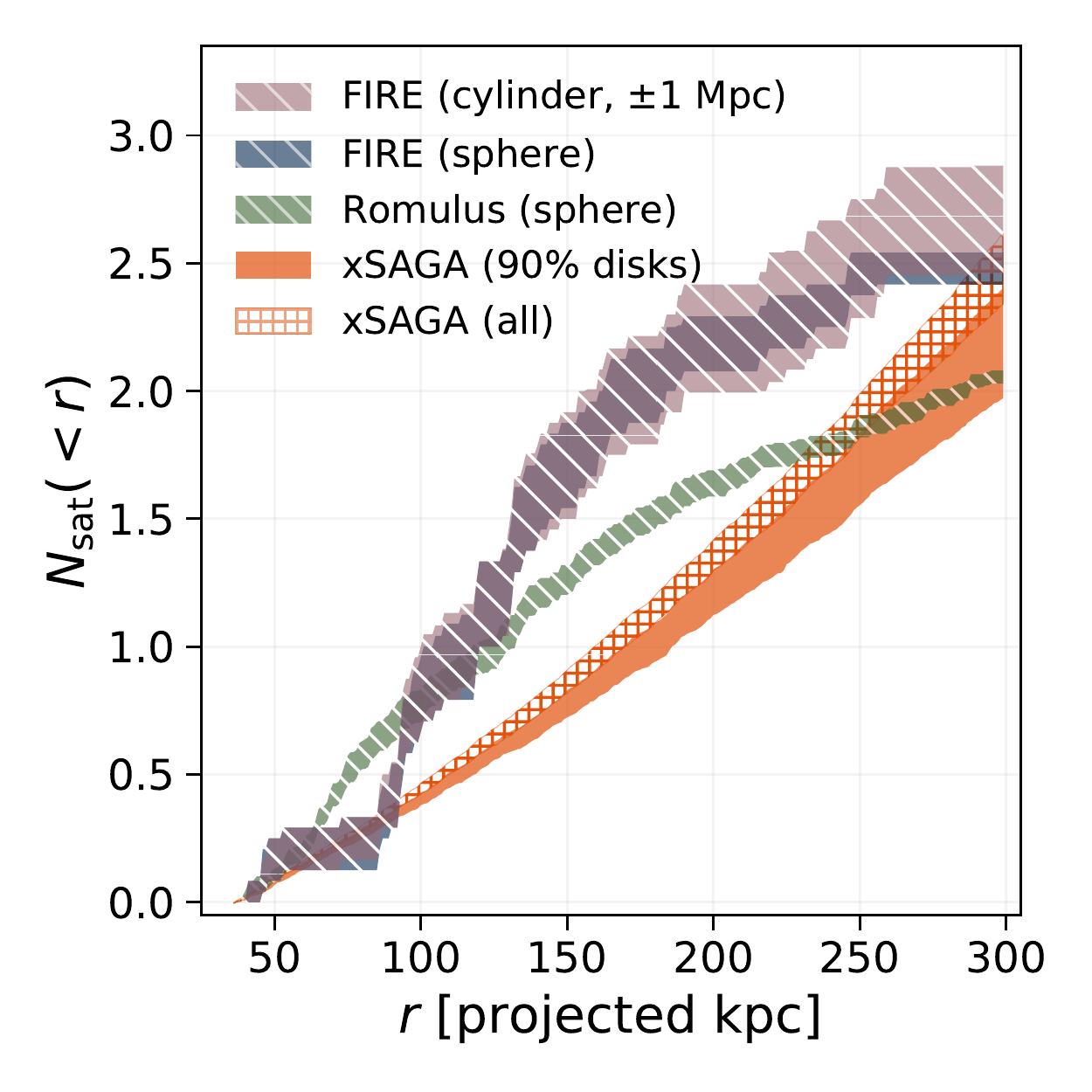}
    \includegraphics[width=0.49\textwidth]{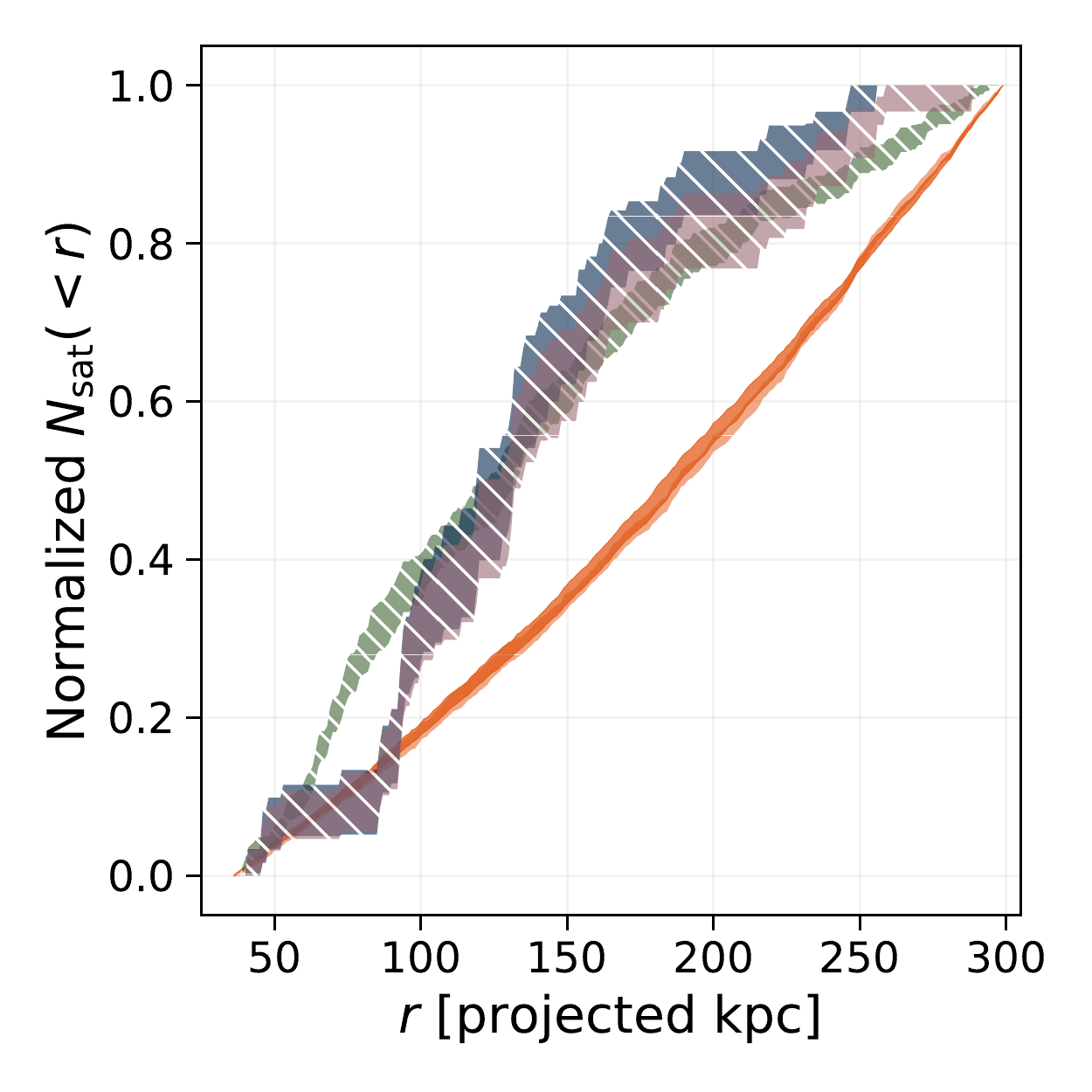}
    \caption{xSAGA satellite radial profiles in projection (\textit{solid orange}) compared with those from the FIRE simulations \citep{Samuel+20} and Romulus25 simulations \citep{Tremmel+17}.
    {We show both unnormalized (\textit{left}) and normalized (\textit{right}) radial profiles.
    We include CNN-selected satellites around a random subset of hosts that have morphologies in proportion to the simulated hosts (90\% disky hosts; \textit{solid orange}) as well as the entire host sample (\textit{hatched orange}).}
    Simulated satellites are selected assuming a spherical geometry with $300$~kpc radius centered on the host for both FIRE (\textit{hatched blue}) and Romulus (\textit{hatched green}), and assuming a cylindrical geometry with line-of-sight distance of $\pm 1$~Mpc centered on the host for FIRE (\textit{hatched lilac}).
    We only compare against simulated satellites with stellar masses greater than $10^{7.5}~M_\odot$.
    }
    \label{fig:compare-sims}
\end{figure*}

\section{Comparison to simulations} \label{sec:sims}

We present an initial comparison to a limited selection of hydrodynamic simulations in order to place the xSAGA results in a theoretical context.
We focus on the satellite richnesses and radial profiles of isolated hosts and how they vary with host galaxy properties.
Although other works have examined large-scale environment, dark matter self-interactions, halo parameters, halo formation and subhalo accretion histories, and their respective impacts on satellite properties \citep[e.g.,][among many others]{Nagai&Kravtsov05,vandenBosch+05,Zentner+05,Diemand+07,Croton+16,Deason+16,Moline+17,Bullock&Boylan-Kolchin17,Buck+19,Robles+19,Wright+19,Richings+20,Green+21a}, we restrict our comparisons to host galaxy stellar mass and morphology in order to facilitate a direct comparison between xSAGA and hydrodynamic simulations, and leave deeper investigations for future analyses.

In Figure~\ref{fig:compare-sims}, we compare the xSAGA radial profiles against results from the FIRE \citep{Samuel+20} and Romulus25 \citep{Tremmel+17} simulations.
All satellite radial profiles are shown in projection and have been bootstrapped.
We select simulated satellites that are located inside a sphere with a 300~kpc radius centered on the host (for both FIRE and Romulus), or inside a cylinder with a 300~kpc projected radius and a $\pm 1$~Mpc line-of-sight distance (FIRE).

The lilac and blue shaded regions show satellite radial profiles from the FIRE simulations, which include all $\log(M_\star/M_\odot) > 7.5$ satellites around eight MW-like hosts from the FIRE simulations.
We select only isolated simulated hosts rather than those in Local Group analogs; however, the overall results do not change if we also include six additional massive hosts in paired configurations when computing FIRE profiles \citep[as expected from][]{Garrison-Kimmel+19}.
FIRE host galaxies range in stellar mass from $10.2 < \log(M_\star/M_\odot) < 11$ (with a mean of $10.7$).
FIRE simulations predict $N_{\rm sat} = 2.50 \pm 0.07$ satellites for a spherical geometry, and $N_{\rm sat} = 2.69 \pm 0.24$ satellites for a cylindrical geometry. 
The Romulus simulation (shown in green) contains 46 MW analogs selected analogous to the SAGA~II host galaxy criteria ($10 < \log(M_\star/M_\odot) < 11$; no galaxy with a $\Delta K < -1.6$ magnitude companion within 700~kpc). 
The satellite abundances for Romulus are significantly lower: $N_{\rm sat} = 2.05 \pm 0.03$.

The number of xSAGA satellites depends strongly on host stellar mass, as we have discussed in Section~\ref{sec:mass}.
The Romulus simulations also find that host stellar mass strongly correlates with satellite richness.
However, \cite{Samuel+20} reported that the number of satellite galaxies within 300~kpc does not significantly vary with host stellar mass (for $M_\star > 10^5~M_\odot$ satellites), and that the satellite abundance is instead dominated by host-to-host scatter.
We note that all of the FIRE hosts and roughly $90\%$ of the Romulus hosts have disky morphologies.
{The left panel of Figure~\ref{fig:compare-sims} shows that, for a matched stellar mass range, the xSAGA satellite richness across all hosts (hatched orange) is consistent with FIRE results: $N_{\rm sat} = 2.56 \pm 0.21$.
When we restrict our comparison to a random subset of 90\% disky hosts in the same stellar mass range (solid orange), we find $N_{\rm sat} = 2.22 \pm 0.26$, which appears to be more in line with the satellite richness for Romulus simulations}.

{In the right panel of Figure~\ref{fig:compare-sims},} we find that the {average projected satellite radial profiles for xSAGA and simulations} appear to have different shapes, but again we note that host-to-host variations dominate the scatter.
The FIRE and SAGA satellite radial profiles are consistent {(\citealt{Samuel+20}; see Figure~14 of \citealt{SAGA-p2})}; since we have shown that the xSAGA radial profile agrees with that of SAGA (Figure~\ref{fig:SAGA-comparison}), then we also expect our results to be in agreement with those from the FIRE simulation.

Finally, we note that \cite{SAGA-p2} compared SAGA~II satellite populations to predictions from an empirical galaxy--halo connection model based on abundance matching \citep{Nadler+19,Nadler+20}. 
The predictions from this model, which has only been fit to MW satellite data, are in reasonable agreement with SAGA~II data, and (by extension, e.g.\ based on Figure~\ref{fig:SAGA-comparison}) with the xSAGA results presented here. 
Future work that combines a generalized version of the \cite{Nadler+20} model with our methodology will allow xSAGA satellites to be interpreted in the context of the galaxy-halo connection while facilitating tests for central galaxy-dependent variations in satellite disruption efficiency.

\section{Discussion} \label{sec:conclusions}

We have introduced xSAGA, a novel machine learning method for identifying $z < 0.03$ (low-$z$) galaxies using Legacy Survey image cutouts.
By using a CNN optimized on a proprietary catalog of galaxy redshifts from the SAGA Survey (SAGA II+ catalog), we are able to identify $\sim 110,000$ probable low-$z$ galaxies.
Our method highlights the power of using CNNs for studying the nearby universe.
CNNs appear to be particularly adept at recognizing low-surface brightness components of low-$z$ galaxies \citep[e.g.,][]{Tanoglidis+21,Martinez-Delgado+21}.
Similarly, CNNs can robustly discover other astronomical phenomena that may be rare but have distinguishing morphological features \citep[e.g.,][]{Stein+21}.
Our work presents an exciting new approach for studying cosmic substructure in the low-redshift Universe.

In this work, we have extended the redshift range of satellite candidates from $z = 0.008$ (SAGA) to $0.01 < z < 0.03$ (xSAGA), and the apparent magnitude limit from $r < 20.75$ to $r < 21.0$.
This difference complicates the selection effects of low-$z$ candidates in a few ways.
First, the limiting absolute magnitude for the CNN-selected sample is fainter ($M_r = -15.0$) compared to that for the SAGA Survey ($M_r = -12.3$; \citealt{SAGA-p1}).
Second, the SAGA~II photometric cuts have been specifically designed to select $z = 0.008$ galaxies, and are already known to be incomplete for more distant galaxies \citep{SAGA-p2}.
Approximately 11\% of $z < 0.03$ galaxies in the SAGA~II+ catalog are outside the SAGA~II photometric cuts because they have too high surface brightness or red $g-r$ color (see Section~\ref{sec:training-sample}).
Finally, very low-surface brightness galaxies are spectroscopically incomplete; \cite{SAGA-p2} have characterized this incompleteness as a function of photometric properties for SAGA satellite candidates.
However, as we have noted above, our CNN is well-suited for finding such objects and will be valuable for targeting the promising diffuse low-$z$ candidate galaxies (i.e., DESI LOWZ Program; E. Darragh-Ford et al., in preparation).

Our CNN selection technique has additional limitations.
The CNN is unable to perfectly distinguish low-$z$ and high-$z$ galaxies, so it suffers from incompleteness and contamination.
We have identified two sources of contamination: (\textit{i}) CNN-misclassified low-$z$ galaxies (FPs) and (\textit{ii}) non-satellite or unrelated low-$z$ galaxies that happen to be projected within 300 kpc of a $\log(M_\star/M_\odot)>9.5$ host galaxy.
We have found that the distribution of FP contaminants are approximately uniform on the sky (Appendix~\ref{app:corrections}), and we have assumed that the non-satellite low-$z$ galaxies selected by the CNN have a constant volume density.
Consequently, very nearby host galaxies, which subtend larger solid angles on the sky, are also subject to greater contamination.
We restrict our analysis of satellite systems to those around hosts in the redshift range $0.01 < z < 0.03$.
If contaminants are clustered with the host galaxy beyond the virial radius, as is expected from hierarchical structure formation theory, then our correction term will systematically underpredict the number of contaminants. 
Conversely, the number of contaminants should be overpredicted in underdense environments. 
However, as we have noted in Section~\ref{sec:corrections}, these corrections are typically modest in size; we expect an average of 0.6 (median of 0.3) contaminants for hosts in our redshift range.

After adopting our $M_r < -15.0$ completeness limit, we have adjusted the satellite number counts using Equation~\ref{eq:corrected-N_sat}, where we have used the SAGA II+ data set to calibrate the correction terms for incompleteness and contamination.
As discussed in Section~\ref{sec:corrections}, these corrections can affect both the radial profile shape and normalization.
Underestimating (overestimating) the contaminant rates will result in too high (low) radial concentrations of satellites.
However, we have found that our CNN selection method is robust and unbiased with satellite-host separation, host stellar mass, and host morphology within the $9.5 < \log(M_\star/M_\odot) < 11$ host mass range (Appendix~\ref{app:analysis-details}).

As a first application of this extremely rich data set, we measure satellite radial profiles to 300~projected kpc for spectroscopically confirmed host galaxies, and report the following results:
\begin{itemize}
    \item The satellite richness $N_{\rm sat}$ depends strongly on host stellar mass (left panel of Figure~\ref{fig:profile}). 
    \item The shape of the radial profile, i.e. the normalized radial distribution of satellites, does not vary with host stellar mass (right panel of Figure~\ref{fig:profile}).
    \item The normalized radial distributions of xSAGA and SAGA satellites are consistent, and their satellite abundances agree to within the bootstrapped standard error (Figure~\ref{fig:SAGA-comparison}).
    \item Elliptical galaxies generally have more satellites than disk galaxies for hosts with $M_\star > 10^{10}~M_\odot$, although lower-mass disks and ellipticals have comparable satellite richness (Figure~\ref{fig:morphology}).
    \item The magnitude gap between the host and brightest satellite $\Delta m_{r,*}$, which probes the halo accretion history, strongly correlates with the number of satellites (e.g., Figures~\ref{fig:satellite-abundance-300kpc} and \ref{fig:satellite-richness_satellite-gap}).
    \item The normalized radial distributions of satellites do not systematically vary with host morphology (Figure~\ref{fig:morphology_radial-dist}) or magnitude gap (Figure~\ref{fig:radial-distribution_magnitude-gap}).
\end{itemize}

Overall, we find strong correlations between the observed satellite richness and various host properties, but detect surprisingly little variation in the radial distributions of satellites.
Satellite number counts and their radial distributions are important constraints on structure formation \citep{Wechsler&Tinker18}.
For much of our analysis, we have imposed a constant minimum radius cut of 36~kpc, which makes it difficult to determine whether the xSAGA radial distributions are self-similar within the virial radius \citep[similar to the dark matter profile; e.g.,][]{NFW,Moore+99,Elahi+09}, or if the concentrations of the satellite radial profiles vary with properties of the central halo and galaxy \citep[e.g.,][]{Bullock+01,Nagai&Kravtsov05,Zhao+09,Mao+15,Nadler+18,Green+21a}.
However, in Section~\ref{sec:in-virial-radius}, we have also limited our analysis to satellite galaxies within $0.26 < r/R_{\rm vir} < 1.0$ of the host galaxy; by restricting the analysis to a constant range in virial radius, we again find that the radial distribution of satellites is self-similar.
More detailed theoretical modeling can illuminate how the radial distribution of xSAGA satellites is expected to depend on halo accretion and tidal destruction \citep[using observational probes such as the magnitude gap; e.g., see also ][]{TothOstriker92,Sales+13,Ostriker+19,Green+21b,Smercina+21,Zhang+21}.

We have compared our results to a limited set of hydrodynamical simulations of satellite galaxy systems.
Following \cite{Samuel+20}, we have exclusively compared un-normalized satellite radial profiles, and found general agreement between xSAGA satellite abundances and the numbers of $M_\star > 10^{7.5}~M_\odot$ satellites in the FIRE and Romulus25 simulations (Figure~\ref{fig:compare-sims}).
These limited comparisons also reveal some differences in the shapes of the satellite radial profiles, although the scatter in these radial profiles is dominated by host-to-host variations.
In order to facilitate more accurate comparisons between theory and observations, we would need to account for the observed geometry (i.e., $2d$ projections for xSAGA), and also include large-scale structure contamination in the mock observations of simulated satellite systems.
Robust accounting for these effects will be critical for measuring any differences in the radial distributions of observed and simulated satellites, and for testing the impacts of baryonic physics on satellite disruption and galaxy evolution \citep[e.g.,][]{Brooks&Zolotov14,Garrison-Kimmel+17,Nadler+18,Kelley+19,Richings+20,Webb&Bovy20,Engler+21,Green+21a,Green+21b}.

Our xSAGA method and CNN-selected sample enable a wide array of new scientific analyses, including (i) a more detailed theory-driven model in order to constrain halo properties, (ii) a comparison of the satellite systems around isolated and paired host galaxies, (iii) an in-depth study of satellite luminosity functions, (iv) an investigation of the full spatial distributions of satellite galaxies, and (v) an analysis of the correlations between satellite system properties and large-scale structure, featuring all low-$z$ galaxies in the xSAGA catalogs.
We aim to address these pursuits in future work.

\acknowledgments
We thank Jenna Samuel for providing data from the Latte suite and ELVIS suite \citep{Wetzel+16,Garrison-Kimmel+19,Samuel+20} of cosmological zoom-in baryonic simulations, which are part of the Feedback In Realistic Environments (FIRE) project, run using the Gizmo code \citep{Hopkins15} and the FIRE-2 physics model \citep{Hopkins+18}.
We thank Jordan van Nest for providing data from the Romulus25 simulations \citep{Tremmel+17}.
We also thank Anna Wright and Azadeh Fattahi for helpful conversations.

Support for YYM was provided by NASA through the NASA Hubble Fellowship grant no.\ HST-HF2-51441.001 awarded by the Space Telescope Science Institute, which is operated by the Association of Universities for Research in Astronomy, Incorporated, under NASA contract NAS5-26555.

This research made use of data from the SAGA Survey (\url{https://sagasurvey.org}). The SAGA Survey was supported by NSF collaborative grants AST-1517148 and AST-1517422 and by Heising–Simons Foundation grant 2019-1402.

Funding for the SDSS and SDSS-II has been provided by the Alfred P. Sloan Foundation, the Participating Institutions, the National Science Foundation, the U.S. Department of Energy, the National Aeronautics and Space Administration, the Japanese Monbukagakusho, the Max Planck Society, and the Higher Education Funding Council for England. The SDSS Web Site is \url{http://www.sdss.org/}.

The SDSS is managed by the Astrophysical Research Consortium for the Participating Institutions. The Participating Institutions are the American Museum of Natural History, Astrophysical Institute Potsdam, University of Basel, University of Cambridge, Case Western Reserve University, University of Chicago, Drexel University, Fermilab, the Institute for Advanced Study, the Japan Participation Group, Johns Hopkins University, the Joint Institute for Nuclear Astrophysics, the Kavli Institute for Particle Astrophysics and Cosmology, the Korean Scientist Group, the Chinese Academy of Sciences (LAMOST), Los Alamos National Laboratory, the Max-Planck-Institute for Astronomy (MPIA), the Max-Planck-Institute for Astrophysics (MPA), New Mexico State University, Ohio State University, University of Pittsburgh, University of Portsmouth, Princeton University, the United States Naval Observatory, and the University of Washington.

The Legacy Surveys consist of three individual and complementary projects: the Dark Energy Camera Legacy Survey (DECaLS; Proposal ID \#2014B-0404; PIs: David Schlegel and Arjun Dey), the Beijing-Arizona Sky Survey (BASS; NOAO Prop. ID \#2015A-0801; PIs: Zhou Xu and Xiaohui Fan), and the Mayall z-band Legacy Survey (MzLS; Prop. ID \#2016A-0453; PI: Arjun Dey). DECaLS, BASS and MzLS together include data obtained, respectively, at the Blanco telescope, Cerro Tololo Inter-American Observatory, NSF's NOIRLab; the Bok telescope, Steward Observatory, University of Arizona; and the Mayall telescope, Kitt Peak National Observatory, NOIRLab. The Legacy Surveys project is honored to be permitted to conduct astronomical research on Iolkam Du'ag (Kitt Peak), a mountain with particular significance to the Tohono O'odham Nation.

NOIRLab is operated by the Association of Universities for Research in Astronomy (AURA) under a cooperative agreement with the National Science Foundation.

This project used data obtained with the Dark Energy Camera (DECam), which was constructed by the Dark Energy Survey (DES) collaboration. Funding for the DES Projects has been provided by the U.S. Department of Energy, the U.S. National Science Foundation, the Ministry of Science and Education of Spain, the Science and Technology Facilities Council of the United Kingdom, the Higher Education Funding Council for England, the National Center for Supercomputing Applications at the University of Illinois at Urbana-Champaign, the Kavli Institute of Cosmological Physics at the University of Chicago, Center for Cosmology and Astro-Particle Physics at the Ohio State University, the Mitchell Institute for Fundamental Physics and Astronomy at Texas A\&M University, Financiadora de Estudos e Projetos, Fundacao Carlos Chagas Filho de Amparo, Financiadora de Estudos e Projetos, Fundacao Carlos Chagas Filho de Amparo a Pesquisa do Estado do Rio de Janeiro, Conselho Nacional de Desenvolvimento Cientifico e Tecnologico and the Ministerio da Ciencia, Tecnologia e Inovacao, the Deutsche Forschungsgemeinschaft and the Collaborating Institutions in the Dark Energy Survey. The Collaborating Institutions are Argonne National Laboratory, the University of California at Santa Cruz, the University of Cambridge, Centro de Investigaciones Energeticas, Medioambientales y Tecnologicas-Madrid, the University of Chicago, University College London, the DES-Brazil Consortium, the University of Edinburgh, the Eidgenossische Technische Hochschule (ETH) Zurich, Fermi National Accelerator Laboratory, the University of Illinois at Urbana-Champaign, the Institut de Ciencies de l’Espai (IEEC/CSIC), the Institut de Fisica d’Altes Energies, Lawrence Berkeley National Laboratory, the Ludwig Maximilians Universitat Munchen and the associated Excellence Cluster Universe, the University of Michigan, NSF’s NOIRLab, the University of Nottingham, the Ohio State University, the University of Pennsylvania, the University of Portsmouth, SLAC National Accelerator Laboratory, Stanford University, the University of Sussex, and Texas A\&M University.

BASS is a key project of the Telescope Access Program (TAP), which has been funded by the National Astronomical Observatories of China, the Chinese Academy of Sciences (the Strategic Priority Research Program “The Emergence of Cosmological Structures” Grant \#XDB09000000), and the Special Fund for Astronomy from the Ministry of Finance. The BASS is also supported by the External Cooperation Program of Chinese Academy of Sciences (Grant \#114A11KYSB20160057), and Chinese National Natural Science Foundation (Grant \#11433005).

The Legacy Survey team makes use of data products from the Near-Earth Object Wide-field Infrared Survey Explorer (NEOWISE), which is a project of the Jet Propulsion Laboratory/California Institute of Technology. NEOWISE is funded by the National Aeronautics and Space Administration.

The Legacy Surveys imaging of the DESI footprint is supported by the Director, Office of Science, Office of High Energy Physics of the U.S. Department of Energy under Contract No. DE-AC02-05CH1123, by the National Energy Research Scientific Computing Center, a DOE Office of Science User Facility under the same contract; and by the U.S. National Science Foundation, Division of Astronomical Sciences under Contract No. AST-0950945 to NOAO.

\software{
    astroML \citep{astroml}, 
    AstroPy \citep{astropy},
    Colossus \cite{COLOSSUS}
    Fastai \citep{fastai},
    matplotlib \citep{matplotlib}, 
    NumPy \citep{numpy},  
    Pytorch \citep{pytorch}
}

\appendix

\section{Convolutional Neural Network details} \label{app:cnn-details}

A neural network model can be trained, or optimized, using a subset of the data known as the training set.
We describe the model architecture in detail in Section~\ref{sec:cnn-arch}.
First, the model ingests ``batches'' of training examples in parallel and outputs predictions for them; this step is known as the forward pass.
We randomly rotate or flip the input images using invariant dihedral group transformations (as in \citealt[][]{Dieleman15}) before passing them into the CNN.
In Section~\ref{sec:cnn-loss}, we evaluate the loss function, which measures the discrepancy between model predictions and labels for each batch of data.
For our binary classification problem, the model predicts the probability that each image is a low-redshift galaxy or a high-redshift interloper (such that these probabilities sum to unity); the labels are either zero or one depending on the example's spectroscopic redshift.

In Section~\ref{app:cnn-opt}, we discuss the details of how model parameters are updated, which is sometimes called the backward pass or update step.
After computing the mean loss for each batch of examples, we calculate contributions to the gradient of the loss for every parameter in the model using the backpropagation algorithm.
We rely on Pytorch to numerically compute these gradients using automatic differentiation.
The model parameters (also known as weights and biases) are updated according to the gradient contributions multiplied by a hyperparameter called the \textit{learning rate}.
A combined forward and backward pass on a single batch of data is called an iteration, and iterating through the entire training data set is called an epoch.
We repeat this training process until the loss converges.

\subsection{Architecture} \label{sec:cnn-arch}

Modern deep neural networks are composed of regular sequences of convolutional layers, feature normalization, and non-linear activation functions \citep[e.g.,][]{Lecun+98,Krizhevsky+12,batchnorm}.
We also include simple self-attention after each convolutional layer in order to help the model encode long-range connections between features \citep{simple-self-attention}.

Typically, batch normalization (batchnorm) layers keep track of the mean and variance of the features after each convolutional layer and empirically improve optimization stability \citep{batchnorm}.
Following \citet{WuPeek2020}, we swap out batchnorm in favor of \textit{deconvolution} in the early layers of the CNN.
Network deconvolution is a form of feature normalization that promotes sparse representation of morphological and color features \citep{deconvolution}.
Later convolutional layers are followed by batchnorm layers as usual, allowing the model to formulate complex features using redundant information.
This hybrid method of feature normalization significantly improves model performance for astronomical computer vision tasks \citep[i.e., predicting galaxies' optical spectra;][]{WuPeek2020}.

After each feature normalization layer, the model applies the Mish activation function \citep{mish}.
In practice, we have found that models with Mish outperform those with Rectified Linear Unit (ReLU) activations \citep[e.g.,][]{Wu2020}.
This is likely because the resulting loss surface with Mish varies more smoothly than ReLU (which has a discontinuity in its derivative).

We combine these convolutional ``blocks'' into a neural network based on the Residual Neural Network (resnet) family of CNNs \citep{resnets}.
We also include several architecture refinements that enhance performance and increase the model's perceptive field \citep{CNNtricks}.
Since we have included \textbf{h}ybrid \textbf{d}econvolution (for feature normalization) and e\textbf{x}tended the \textbf{resnet} architecture with these other tweaks, we refer to the model as \textbf{hdxresnet} in the text (i.e., \texttt{hdxresnet34} for a 34-layer CNN).
We also compare against a more typical deep neural network (\texttt{resnet34}) optimized using Adam with a BCE loss (see Section~\ref{sec:crossvalidation} for results).
Model parameters are drawn from random samples according to Kaiming initialization \citep{kaiming-init}.

\subsection{Loss function} \label{sec:cnn-loss}

During the forward pass, the CNN layers sequentially operate on the input data.
For classification problems, a softmax function $\sigma(\mathbf{z})_i = \exp(z_i) / \left(\sum_{j=1}^K \exp(z_j)\right)$ converts raw model outputs $z$ into class prediction likelihoods $p$, where $i$ denotes the example and $j$ represents the class. 
Note that we have previously used $p_{\rm CNN}$ to present the predictions, but here we will use $p$ with no ambiguity.
Predictions can be compared against the class labels, $y$, which are often represented as unit vectors in the $K$-dimensional space of different classes.
A standard choice of loss function is the binary cross entropy (BCE),
\begin{equation} \label{eq:nll}
    \mathrm{BCE}(p, y) =
    \begin{cases}
        w_1\log(p)   & \text{if~} y = 1, \\
        w_0\log(1-p) & \text{if~} y = 0,
    \end{cases}
\end{equation}
where $p \in [0, 1]$ is a single model prediction, $y$ is the binary ground truth label, and $w_0$ and $w_1$ are weight terms.

Our classification task suffers from class imbalance, which can make it difficult to consistently predict the underrepresented class.
Because $98\%$ of the examples are \textit{not} at low redshift (i.e., they have class labels $y=0$), then a model can conceivably achieve 98\% accuracy by always predicting this majority class.
We can try to avoid this behavior by downweighting the negative class labels, but in practice this only results in modest gains.
Another option is to over-sample the $y=1$ class, but this decreases the purity of the predicted low-$z$ galaxy sample, while increasing its completeness.
A third possibility is to use ``label-smoothing'' cross entropy by adding some noise $\epsilon$ to the ground truth labels; label-smoothing prevents the model from becoming too confident in any of its predictions \citep[e.g.,][]{LabelSmoothing}.

A fourth idea is to modify the cross entropy loss in a way such that the optimizer is forced to hone in on difficult examples while not spending much time on easy examples.
This function, focal loss \citep[FL;][]{FocalLoss}, was initially proposed to solve the imbalanced task of classifying pixels in images as foreground objects of interest, as opposed to background pixels, but it performs extremely well in other imbalanced classification problems.
The focal loss is defined:

\begin{equation} \label{eq:focalloss}
    \mathrm{FL}(p, y) =
    \begin{cases}
        -(1-p)^\gamma \log(p) & \text{if~} y = 1,  \\
        - p^\gamma \log(1-p)  & \text{if~} y = 0,
    \end{cases} 
\end{equation}
which contains a hyperparameter $\gamma$ that controls how much the optimizer focuses on challenging examples.
Recent work has also shown that FL improves model calibration and assists with generalization because it effectively regularizes the BCE loss via maximizing the entropy of the distribution of predictions \citep[e.g.,][]{FL-calibration}.
Our tests and previous literature indicate that $\gamma = 2$ is a good choice for this hyperparameter.

\subsection{Optimization} \label{app:cnn-opt}

An optimizer is an algorithm designed to update model parameters in a way that minimizes the loss.
Efficient optimizers are able to guide the model toward regions of low loss in a small number of iterations.
The most simple algorithm is stochastic gradient descent (SGD), which computes the partial derivative of the loss (given a batch of training data) with every parameter in the neural network model.
In other words, SGD finds the update direction that would result in steepest descent of the loss.
The model parameters then take a step in the direction computed by the optimizer, with a step size dictated by the learning rate.
Usually the learning rate is small ($\ll 1$) in order to keep the loss stable.
After many training epochs, the optimizer should converge on a low loss. 

SGD can be quite inefficient because each update has no memory of previous updates.
In practice, SGD will direct model updates toward approximately the same direction for many iterations; this is again because the learning rate is usually small.
An alternative optimizer, Adam, is popular because it can achieve quicker convergence and can be generically applied to many machine learning problems \citep{Adam}. 
Different optimizers have dynamics similar to those of physical systems \citep[e.g.,][]{Goh17}; Adam adds momentum and friction terms to SGD in order to speed up its traversal of the loss landscape.
We use a optimizer called Ranger that offers some additional improvements over Adam \citep{Ranger} and has been shown to yield superior results for astronomical data sets \citep[i.e.,][]{Wu2020}.
A related method for improving the optimization procedure is varying the learning rate over the course of training.
We use the one-cycle schedule introduced in \citep{1cycle}, with a maximum learning rate of $0.01$ and 10 epochs of training.
The CNN quickly achieves convergence in 10 epochs due to our highly optimized training regiment (see also \citealt{WuBoada19} for details). 
Tests with 15 and 20 epochs of training yielded worse cross-validation results, which indicates that the CNN was overfitting the training data.

\section{Details on the analysis} \label{app:analysis-details}

\subsection{Modeling low-$z$ rate, purity, and completeness of individual galaxies} \label{app:modeling-metrics}

We can model the rate of finding a low-$z$ galaxy as a function of its photometric properties, where the low-$z$ rate is defined as the ratio of the number of low-$z$ galaxies to all SAGA targets.
This method is derived from the $\mathcal R_{\rm sat}$ model in \cite{SAGA-p2}, although we do not use it to directly estimate the rates of finding \textit{satellite} galaxies.
We can similarly estimate the CNN purity and completeness as a function of photometric properties.
The benefit of this approach is that we can compute model purity and completeness for individual galaxy candidates, whereas directly computing purity and completeness metrics necessitates binning data into (large) subsamples.
Following \cite{SAGA-p2}, we fit a continuous model to the low-$z$ rate, purity, and completeness as a function of apparent magnitude, surface brightness, and color. 
The model is similar to a logistic function parameterized by latent variable $\ell$.
The latent variable is linear in each of the photometric properties,

\begin{equation}
    \ell = \beta_0 + \beta_1 r_0 + \beta_2 \mu_{r, \rm eff} + \beta_3 (g-r)_0,
\end{equation}
and parameterizes the model (along with parameter $\mathcal R_{\rm max}$),

\begin{equation}
    \rho = \frac{\mathcal R_{\rm max}}{1 + \exp(-\ell)},
\end{equation}
where we model one of $\rho \in $\{rate, completeness, purity\}.

We compute the log-likelihood given by 

\begin{equation}
    \log \mathcal L = \sum_{i\in{\rm low}-z} \log \rho_i + \sum_{i\notin{\rm low}-z}\log \left (1-\rho_i \right),
\end{equation}
and impose a uniform prior between $(-\infty, \infty)$ for all parameters aside from $\mathcal R_{\rm max}$, which we assume is uniformly distributed between $[0, 1]$.
We minimize the negative log posterior using the \texttt{scipy.optimize.minimize} routine.
This process converges on a solution for each model statistic (low-$z$ rate, completeness, and purity).

\begin{figure*}
    \centering
    \plotone{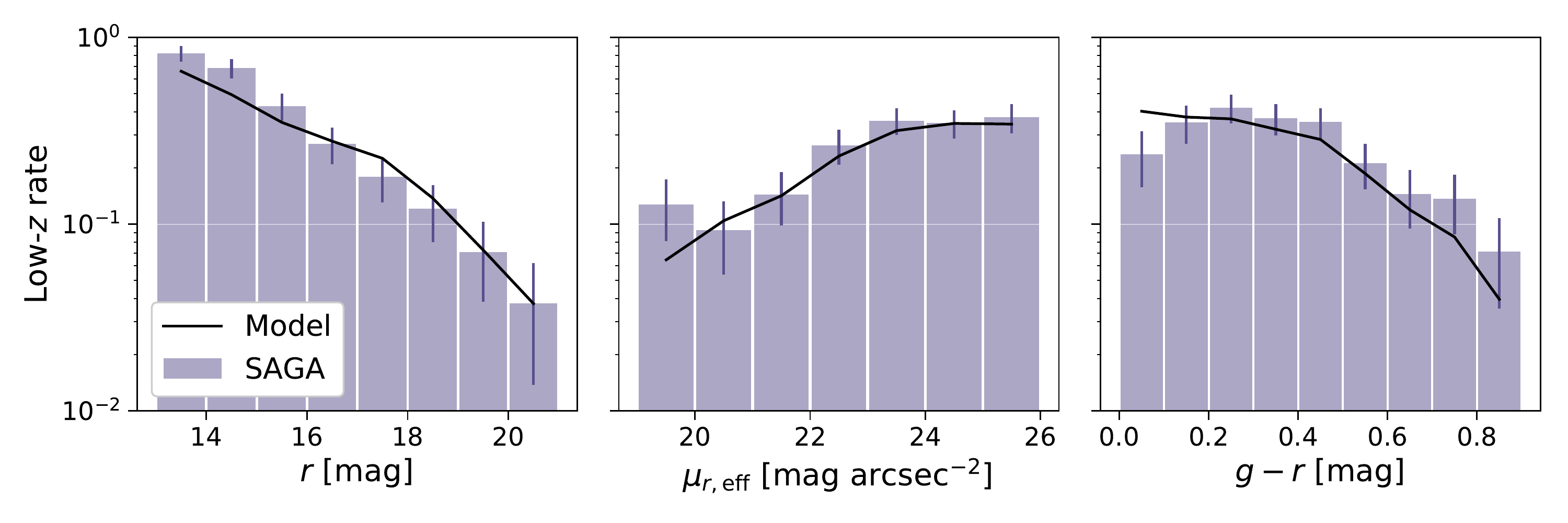}
    \plotone{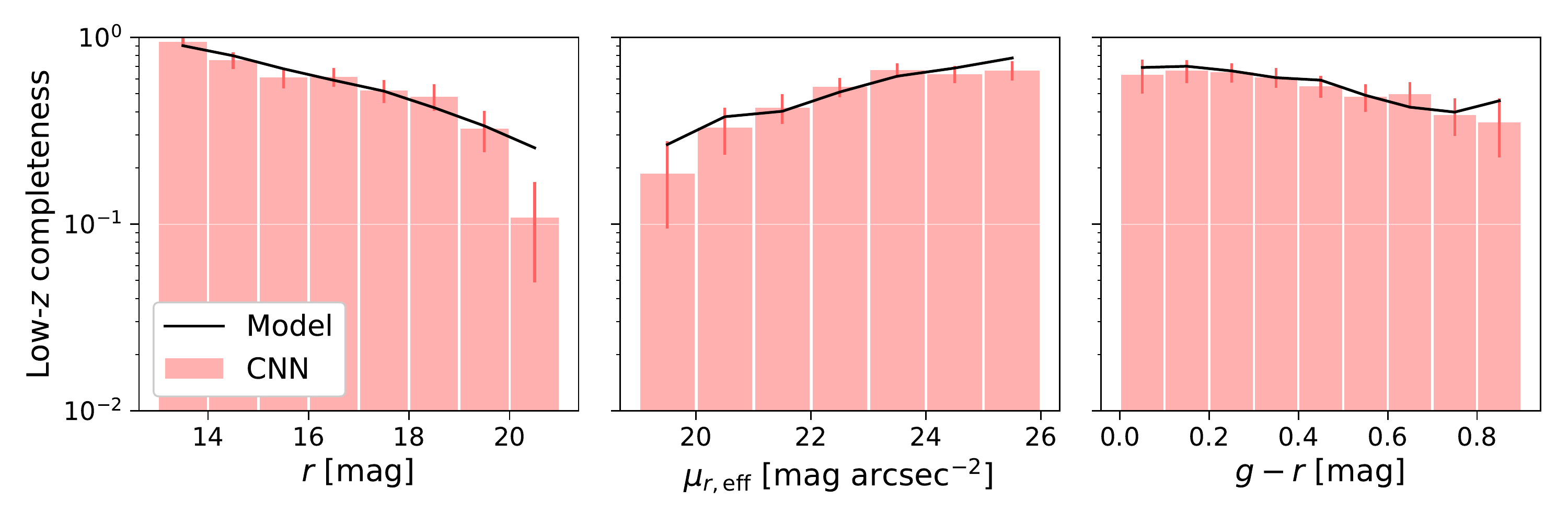}
    \plotone{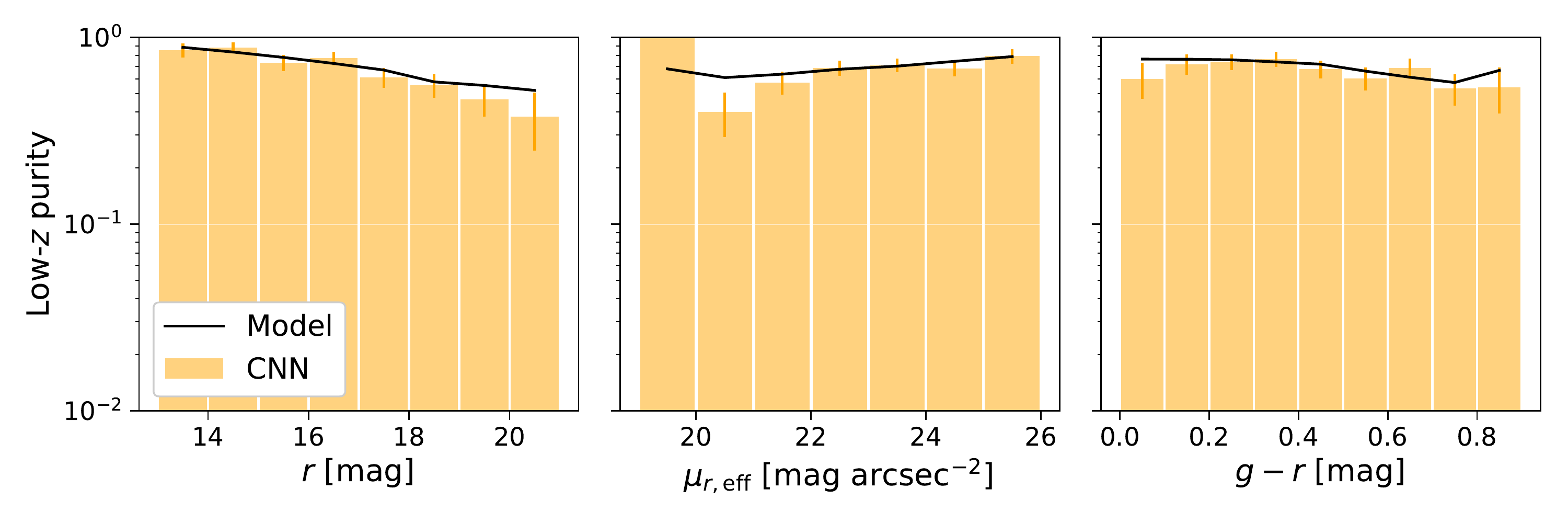}
    \caption{Photometric model fits to the SAGA low-$z$ rate (\textit{top}), CNN completeness (\textit{center}), and CNN purity (\textit{bottom}) using xSAGA cross-validated predictions.
    Data from spectroscopy and CNN predictions (colored histograms) are compared to independent best-fit models (black lines) as a function of magnitude (\textit{left}), surface brightness (\textit{middle}), and color (\textit{right}).
    1$\sigma$ data uncertainties are estimated using the normal approximation for a binomial distribution.
    }
    \label{fig:model-stastics}
\end{figure*}

Figure~\ref{fig:model-stastics} shows our model-fit results alongside the cross-validated data, revealing that our simple model is able to fit the data.
Although the data are binned in the figure for clarity, the models are fit to each of the 112k SAGA training examples.
In terms of completeness and purity, the model performs best in regions of photometric space where there are many examples, but deviates from the ground truths for faint and high-surface brightness objects (where high-$z$ interlopers outnumber true low-$z$ systems).
We may be able to choose a latent variable that is quadratic or higher-order in the photometric variables in order to improve our model, but the linear dependence captures most of the variance and is sufficient for characterizing and correcting CNN errors.
The CNN completeness and purity are much higher than the low-$z$ rate, which highlights our machine learning algorithm's success at identifying probable low-$z$ galaxies.

We note that the empirical purity and completeness can be used to correct the number of FPs and FNs (see Section~\ref{app:corrections} for discussion on the optimal correction method).
We will also now approximate $\mathcal P_{\rm model}/\mathcal C_{\rm model}$ for individual galaxy candidates using our photometric model.
$\mathcal P_{\rm model}/\mathcal C_{\rm model}$ can be used to diagnose trends or biases in the expected purity and completeness of SAGA and xSAGA candidates.
The modeled correction factor is inadequate for estimating corrections to the satellite number counts because high-$z$ galaxies vastly outnumber low-$z$ galaxies, and would result in extremely large uncertainties.
In the next subsection below, we examine whether $\mathcal P_{\rm model}/\mathcal C_{\rm model}$ varies with host galaxy properties. 
If we see no model-estimated trends in purity and completeness, then we can be confident that the CNN selection does not bias our results.

\subsection{Satellite biases with host properties} \label{app:biases}

\begin{figure*}
    \centering
    \plotone{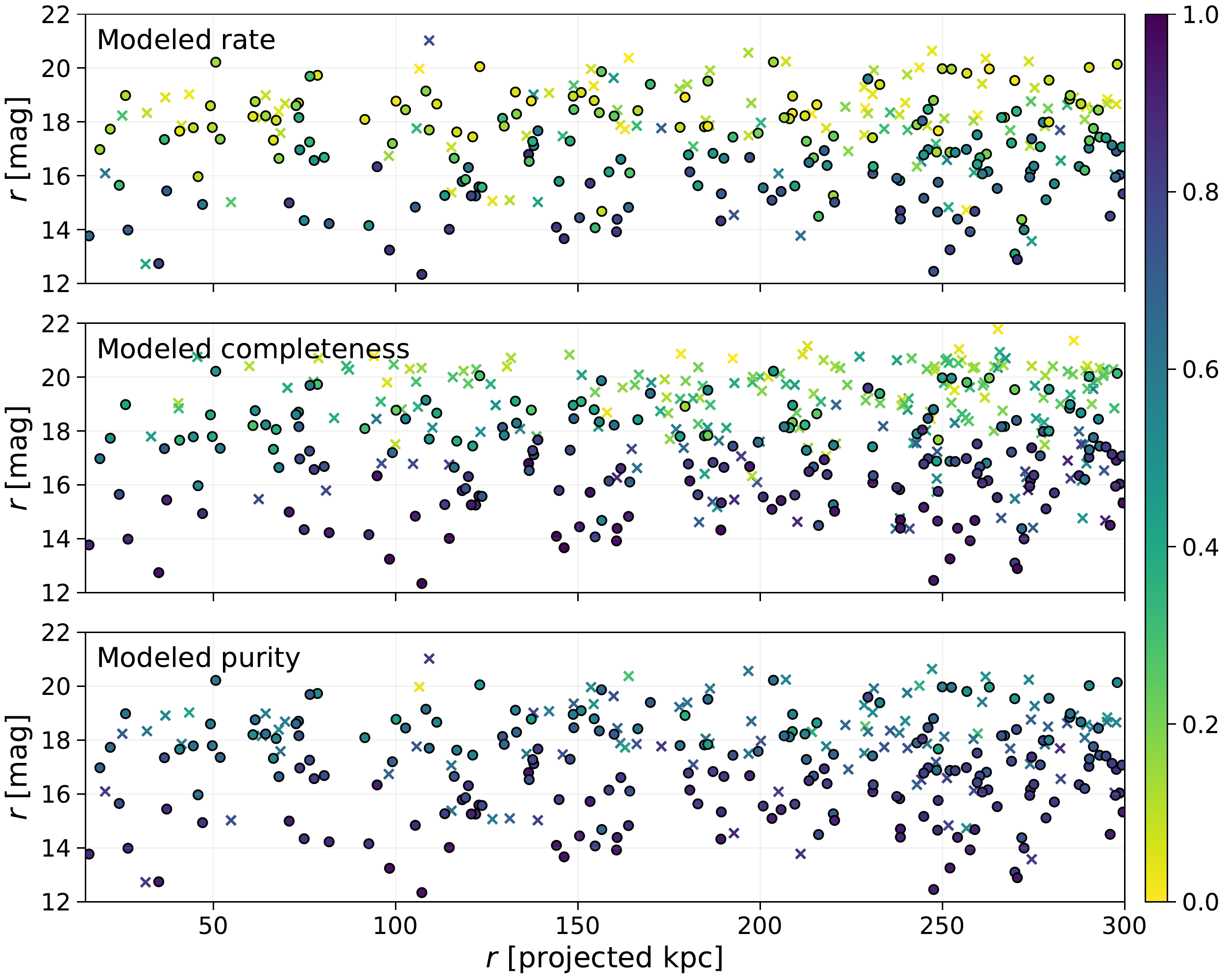}
    \caption{Scatter plots depicting the magnitude versus projected separation for cross-validated low-$z$ objects in SAGA fields. 
    The panels are colored by model predictions for low-$z$ rate (\textit{top}), completeness (\textit{middle}), and purity (\textit{bottom}).
    TP detections from the CNN are shown as circles, while FNs are shown as crosses in the panel showing completeness, and FPs are shown as crosses in the panels showing low-$z$ rate and purity.
    }
    \label{fig:model-metrics-trend}
\end{figure*}

In Figure~\ref{fig:model-metrics-trend}, we show the modeled low-$z$ rate, completeness, and purity for SAGA cross-validation data.
The CNN's completeness and purity both decrease for galaxies at larger separations from the host galaxy.
Such a signal may be due to a bias in the CNN or a radial trend in the photometric properties of SAGA candidates.
Additionally, satellites are more often found close to their hosts, so the completeness and purity statistics are expected to drop at larger separations.
We apply our model to examine if a completeness and purity trend is predicted by the photometric model, or if we need to invoke an additional correction as a function of projected radius.

\begin{figure}
    \centering
    \includegraphics[width=\columnwidth]{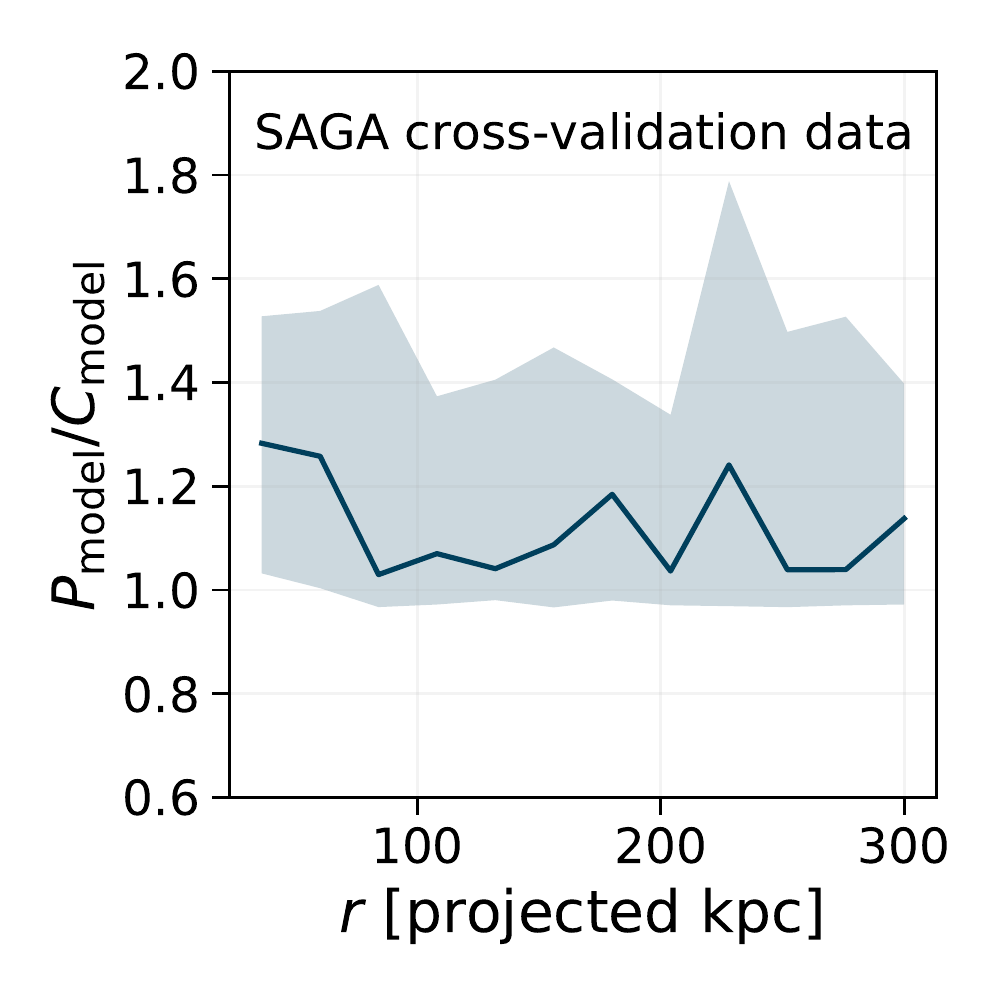}
    \caption{Modeled purity-to-completeness ratio of satellite candidates shown against projected host separation for the SAGA spectroscopically confirmed sample.
    Both the median trend \textit{(solid line}) as well as the 16th through 84th percentile scatter (\textit{shaded region}) are plotted here.
    $\mathcal P_{\rm model} / \mathcal C_{\rm model}$ is centered around a median value of $\sim 1.1$, which is consistent with our earlier findings (i.e., the green histogram in Figure~\ref{fig:TP-corrections}).
    }
    \label{fig:model-PC-with-separation}
\end{figure}

Our photometric model reveals that candidates at larger separations from a host galaxy have somewhat lower completeness and purity.
Part of this effect is driven by the larger area encompassed in a larger radius, which increases the likelihood of finding a galaxy with low model completeness or purity.
Thus, the radial trends in $\mathcal P_{\rm model}$ and $\mathcal C_{\rm model}$ are expected.
Using only the model for low-$z$ \textit{rate} does not reveal such a trend, which indicates that the CNN selection is more robust than using a photometric model alone.
Most importantly, the modeled correction factor $\mathcal P_{\rm model} / \mathcal C_{\rm model}$ stays constant within the $68\%$ scatter for all projected separations (see Figure~\ref{fig:model-PC-with-separation}).
We conclude that that the radial trends in completeness and purity arise solely from the low-$z$ candidate's photometric properties (rather than a CNN bias with radius) and can be corrected using Equation~\ref{eq:corrected-N_sat} above.

\begin{figure*}
    \centering
    \plottwo{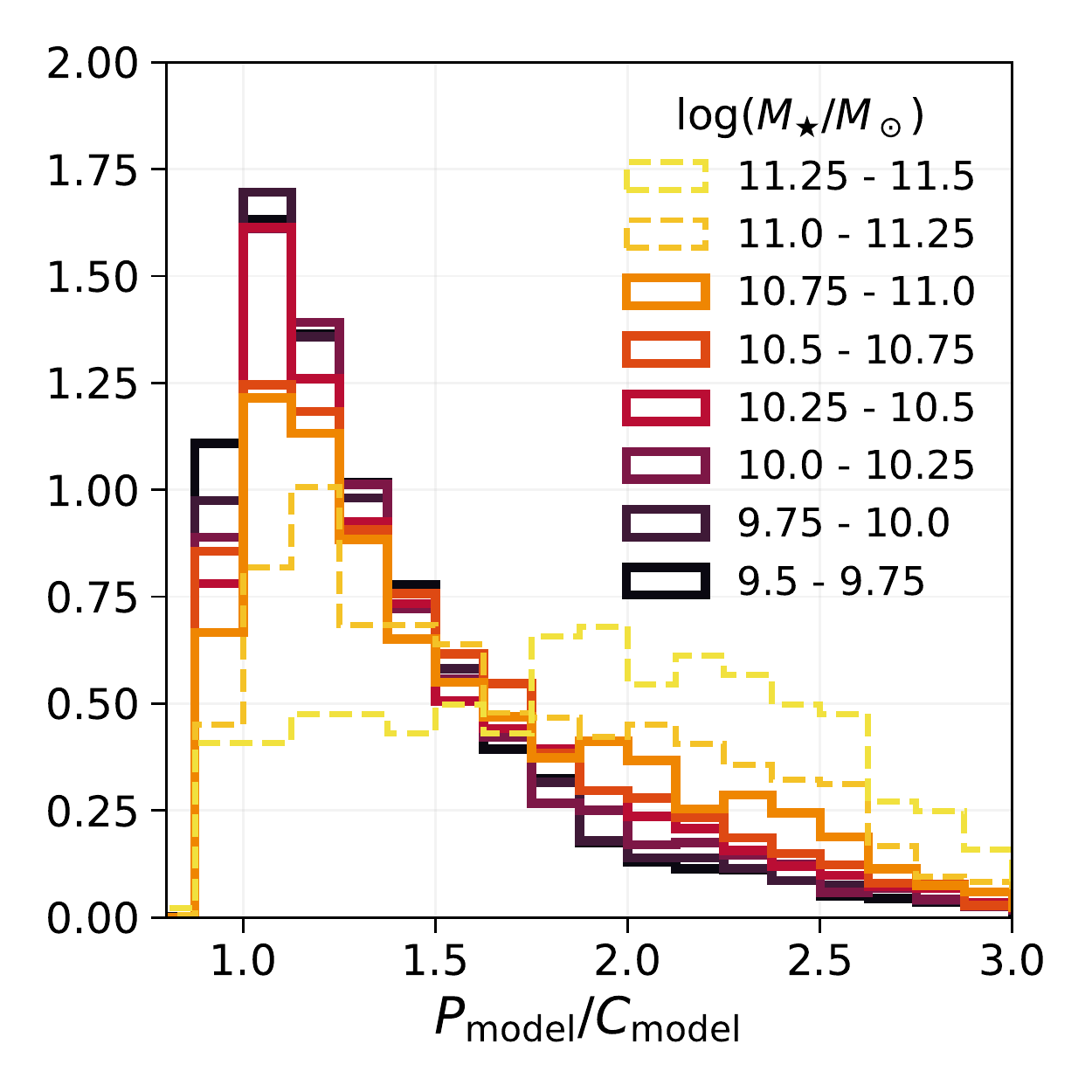}{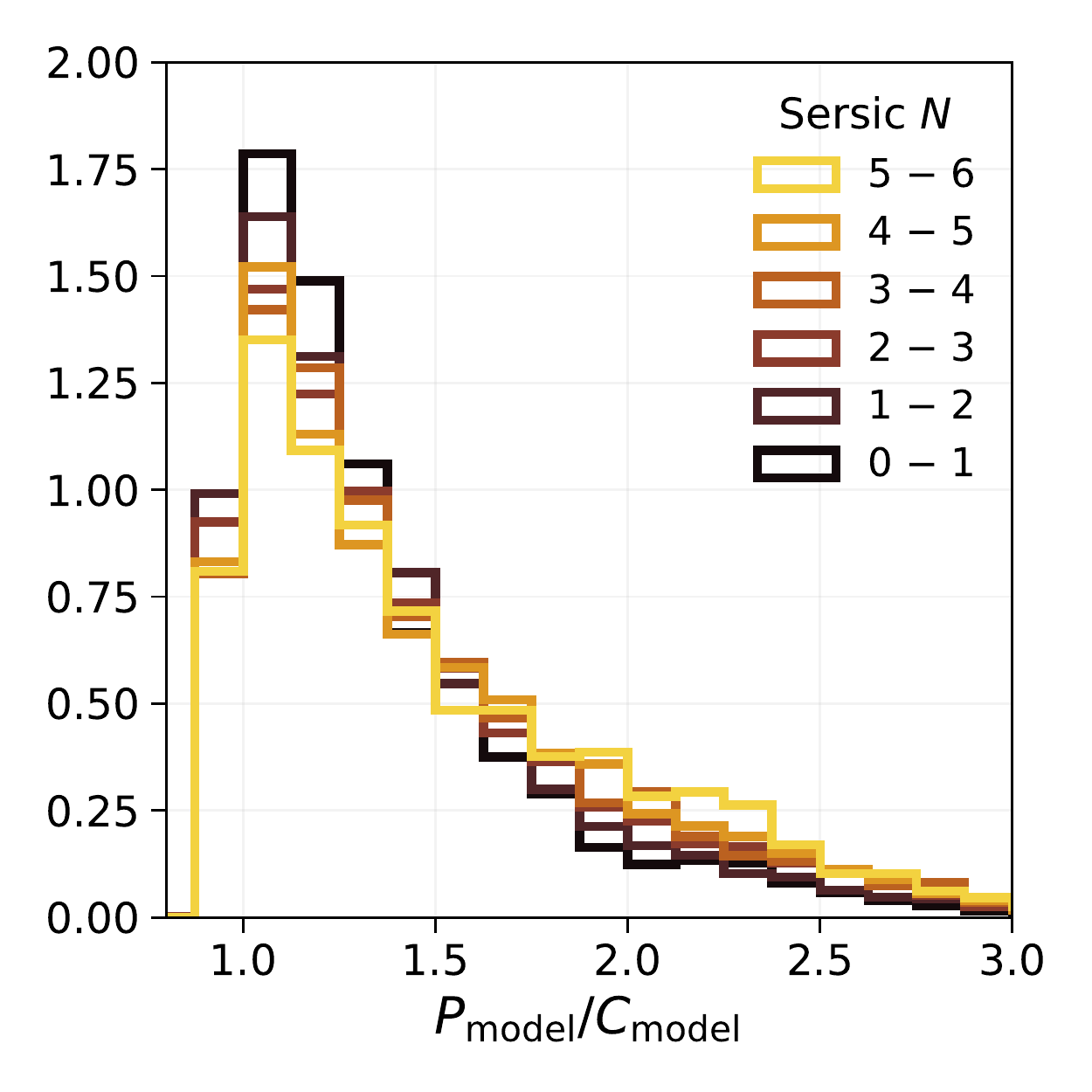}
    \caption{Distributions of modeled purity-to-completeness ratio, $\mathcal P_{\rm model} / \mathcal C_{\rm model}$, aggregated by host stellar mass (\textit{left}) and host Sersic index (\textit{right}) for the xSAGA CNN-selected galaxy sample.
    The left panel shows that the $\mathcal P_{\rm model} / \mathcal C_{\rm model}$ distribution becomes distorted for CNN-selected galaxies around massive hosts ($\log(M_\star / M_\odot) > 11$, depicted with dashed lines).
    Consequently, these satellite systems are removed from our analysis.
    The right panel shows that the $\mathcal P_{\rm model} / \mathcal C_{\rm model}$ distribution is largely insensitive to the Sersic index of the host.
    }
    \label{fig:model-PC-with-host-properties}
\end{figure*}

We also examine whether host galaxy properties such as mass and morphology affect the purity and completeness of xSAGA predictions.
Although we do not have ground-truth (spectroscopic) classifications of whether these objects are at low-$z$ or high-$z$, we can still examine whether the photometric model anticipates trends in the purity and completeness of the CNN-selected sample.
After we assign CNN-selected objects to host galaxies using the procedure described in Section~\ref{sec:assigning-satellites}, we compute $\mathcal P_{\rm model}$ and $\mathcal C_{\rm model}$ for each host galaxy, and compare these metrics as a function of host stellar mass and Sersic index.
These results are shown in Figure~\ref{fig:model-PC-with-host-properties}.

We find that $\mathcal P_{\rm model}$ and $\mathcal C_{\rm model}$ exhibit modest variations with host mass in the range $10^{9.5} - 10^{11}~M_\odot$, but that the ratio between them does not vary.
In other words, the distributions of $\mathcal P_{\rm model}$ and $\mathcal C_{\rm model}$, averaged over each host, become distorted as we vary host stellar mass. 
Nevertheless, the distribution of the ratio of purity and completeness, $\mathcal P_{\rm model}/\mathcal C_{\rm model}$, remains constant over $9.5 < \log(M_\star/M_\odot) < 11$.
We find that $\mathcal P / \mathcal C$ is an effective correction factor for CNN predictions (Section~\ref{app:corrections}).
Therefore, the host-averaged satellite richness can be robustly computed over this range of stellar masses.
However, the correction breaks down for the highest-stellar mass hosts: above $10^{11}~M_\odot$, both the purity and completeness decrease significantly, and $\mathcal P_{\rm model} / \mathcal C_{\rm model}$ also shifts to higher values (see left panel of Figure~\ref{fig:model-PC-with-host-properties}).
The breakdown occurs because many xSAGA hosts in this mass range reside in overdense environments; it is likely that low-$z$ candidates in rich groups or clusters have different properties than those around more field-galaxy hosts.
Ram pressure and tidal stripping effects on satellites are already well-studied for these massive systems \citep[e.g.,][]{Chung+09,Brown+17,Engler+21}, and overdense environments are known to perturb galaxy morphologies in a way that prevents robust CNN selection \citep{Wu2020}. 
Consequently, we exclude low-$z$ galaxies around very massive hosts ($M_\star > 10^{11}~M_\odot$) from our analysis.

We turn our attention to host galaxies morphological parameters, such as the Sersic index and axial ratio \citep[catalogued in NSA;][]{NSA}.
For our CNN-selected sample, we find little variance in modeled purity or completeness with Sersic index between values of $1-6$, and the ratio $\mathcal P_{\rm model} / \mathcal C_{\rm model}$ does not change over this range (see right panel of Figure~\ref{fig:model-PC-with-host-properties}).
We similarly do not see any trends with the host galaxy's axial ratio.
Thus, we conclude that CNN correction factors do not depend on host mass or Sersic index for low-$z$ galaxies for the range of hosts we study.
We proceed using Equation~\ref{eq:corrected-N_sat} to correct the number counts of xSAGA satellites.

\subsection{Ensuring that host galaxies are isolated centrals} \label{app:central-criterion}
Our algorithm for satellite-host assignment may be prone to errors in the event of multiple halos at different redshifts along the line of sight; however, the chances of such an alignment are rare in the $0.01 < z < 0.03$ range. 
Due to the clustering of real halos, the probability that two unrelated hosts have overlapping halos is lower than that of random halos, and much lower than the probability of overlap for two halos at comparable redshifts.
We empirically compute the fraction of overlapping halos at unrelated and similar redshifts for real and randomly generated host locations distributed with constant areal density in the contiguous patch of sky between RA of $120-240$ deg and Dec of $0-50$ deg, assuming a typical halo size of 300 kpc.
The fraction of real overlapping halos at unrelated redshifts is $0.023$, whereas the fraction of overlapping halos at similar redshifts is $0.216$.
The fraction of any two random hosts overlapping is $0.069$.

We enforce a ``isolated'' host criterion that removes galaxies with more massive neighbors within $\Delta z = 0.005$ and $1$~projected Mpc. 
Varying the latter distance threshold between $0.3$ to $2$~projected Mpc has little effect on our results, so we we proceed with this ``isolated'' criterion in addition to the correction terms discussed above.
In comparison, the SAGA Survey identifies host galaxies that are at least 1.6 magnitudes brighter in the $K$-band than any other neighbor within 300~projected kpc (\citealt{SAGA-p2}).

\subsection{Corrections to low-$z$ galaxy number counts} \label{app:corrections}

There are several ways to account for the completeness and impurity of our CNN-selected sample.
The true number of low-$z$ galaxies is equivalent to the number of TPs + FNs, while our CNN identifies $N_{\rm pred}$ galaxies, equal to the number of TPs + FPs.
One possible way to correct the low-$z$ number counts is by estimating $N_{\rm lowz} = \langle \mathcal P / \mathcal C \rangle N_{\rm pred}$.
This method assumes that an average completeness and purity correction factor can be applied to the entire sample.
Indeed, it is standard to compute and divide by $\langle \mathcal C \rangle$ in order to correct for incompleteness.
However, a constant purity factor may not adequately correct for the incidence of FPs.

\begin{figure}
    \centering
    \includegraphics[width=\columnwidth]{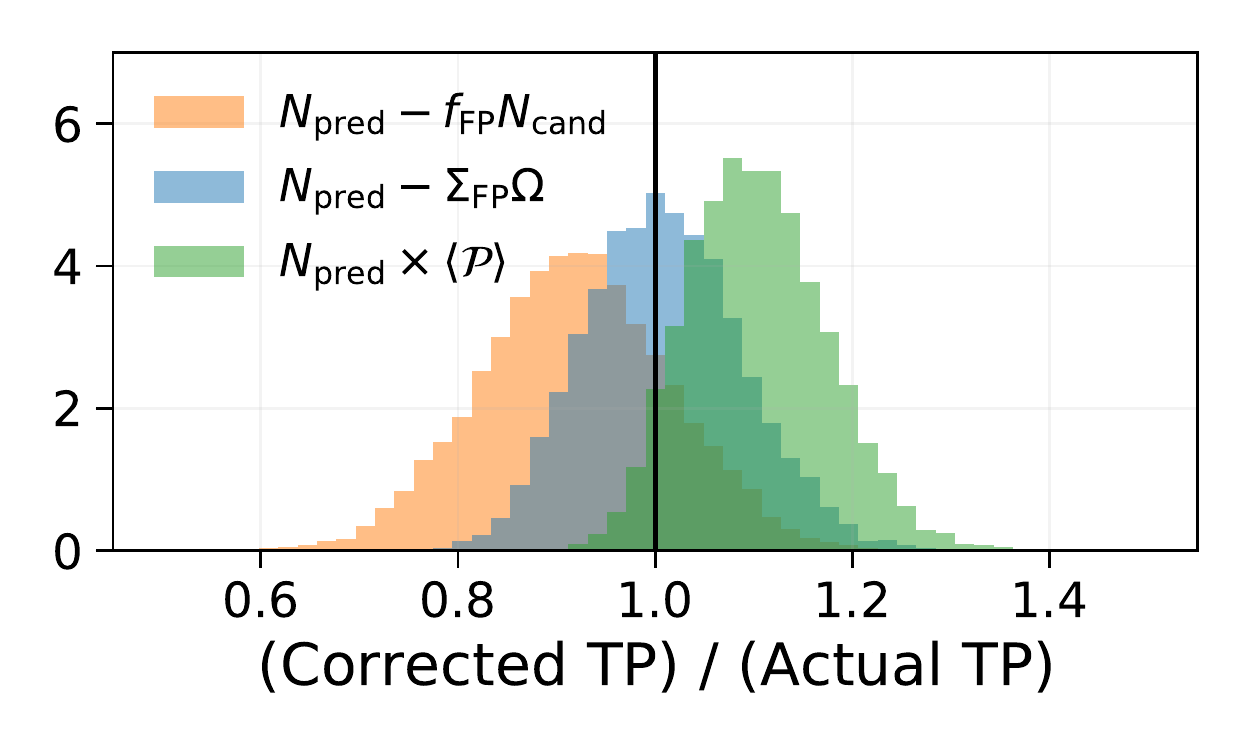}
    \caption{Comparison of adjusted estimates for the number of CNN-identified low-$z$ galaxies (TPs) in SAGA cross validation data. Adjustments are made by either subtracting or multiplying by a correction factor that accounts for FP detections. We correct the number counts by subtracting a mean surface density of FPs (blue), subtracting a mean FP rate for all candidates (orange), and multiplying by a mean purity factor (green). All number counts are normalized by the known number of TPs, such that a value of 1 is optimal. We have bootstrap resampled TP number counts for the SAGA hosts 10,000 times, and computed the mean, standard deviation, and median of the distributions.
    Because the mean FP surface density correction most accurately recovers the number of TPs, we adopt this method for the main analysis.}
    \label{fig:TP-corrections}
\end{figure}

There are other options for removing contaminants based on the number of total candidates or the surveyed area, i.e., by assuming a constant FP rate or a constant FP surface density.
In these cases, the estimated number of contaminants scales with the total number of candidates.
Using SAGA cross-validation, we compute the mean FP rate per SAGA candidate, $\langle f_{\rm FP}\rangle = 1.61 \times 10^{-2}$, as well as the mean FP surface density, $\langle \Sigma_{\rm FP} \rangle = 3.04$~deg$^{-2}$.
We can then adjust the number of low-$z$ galaxies by \textit{subtracting} the average number of FP objects depending on total number of candidates or area.
We can also apply these same corrections to satellite number counts for individual host galaxies.
Using the three methods described above, we compare corrected numbers of low-$z$ objects to the true low-$z$ number counts for each SAGA field in Figure~\ref{fig:TP-corrections}.

We find that the three estimators all provide low-$z$ number counts that are within $\sim 10\%$ of the true value. 
The most robust method is by subtracting a number of FPs that depends on the area, which recovers $1.01 \pm 0.08$ times the correct number of true positives (i.e., in excellent agreement with unity).
The FP rate estimator is prone to slightly under-predicting the number of TPs by a factor of $0.92 \pm 0.09$.
The constant CNN purity correction factor tends to over-predict the number of TPs by a factor of $1.10 \pm 0.07$.
Therefore, we correct for contaminants by subtracting the average surface density of FPs scaled by the surface area. 
The full corrected number of low-$z$ galaxies (per host) is

\begin{equation}\label{eq:corrected-lowz}
    N_{\rm lowz} = \frac{N_{\rm pred}}{\langle \mathcal C\rangle} - \left\langle \Sigma_{\rm FP}\right \rangle \Omega,
\end{equation}
where $\Omega$ is the surface area (of the subtended halo) in square degrees.

We consider an additional correction term that depends on the proper volume of each host galaxy's halo. 
Our CNN method selects galaxies that are at $z < 0.03$ but possibly at a different redshift than the host galaxy.
We issue a correction term for these non-satellite low-$z$ galaxies in Equation~\ref{eq:corrected-N_sat}.
We estimate the volume density of contaminated low-$z$ galaxies using the sample of $0.02 < z < 0.03$ objects in the SAGA II+ catalog, which should be unrelated to SAGA systems at $z \sim 0.008$.
The expected number of contaminants can be computed by assuming a constant volume density of unrelated low-$z$ galaxies in SAGA fields, $\langle n_{\rm nonsat}\rangle = 1.42 \times 10^{-2}~{\rm Mpc}^{-3}$, multiplied by the proper volume $V$ corresponding to the pencil beam geometry around each xSAGA host and outside each halo's redshift range.
We adopt $\Delta z = 0.005$ in order to compute this volume.
All of the correction terms are illustrated in Figure~\ref{fig:method}.

\section{Bootstrapped radial profile} \label{app:scatter}

\begin{figure*}
    \centering
    \includegraphics[width=\textwidth]{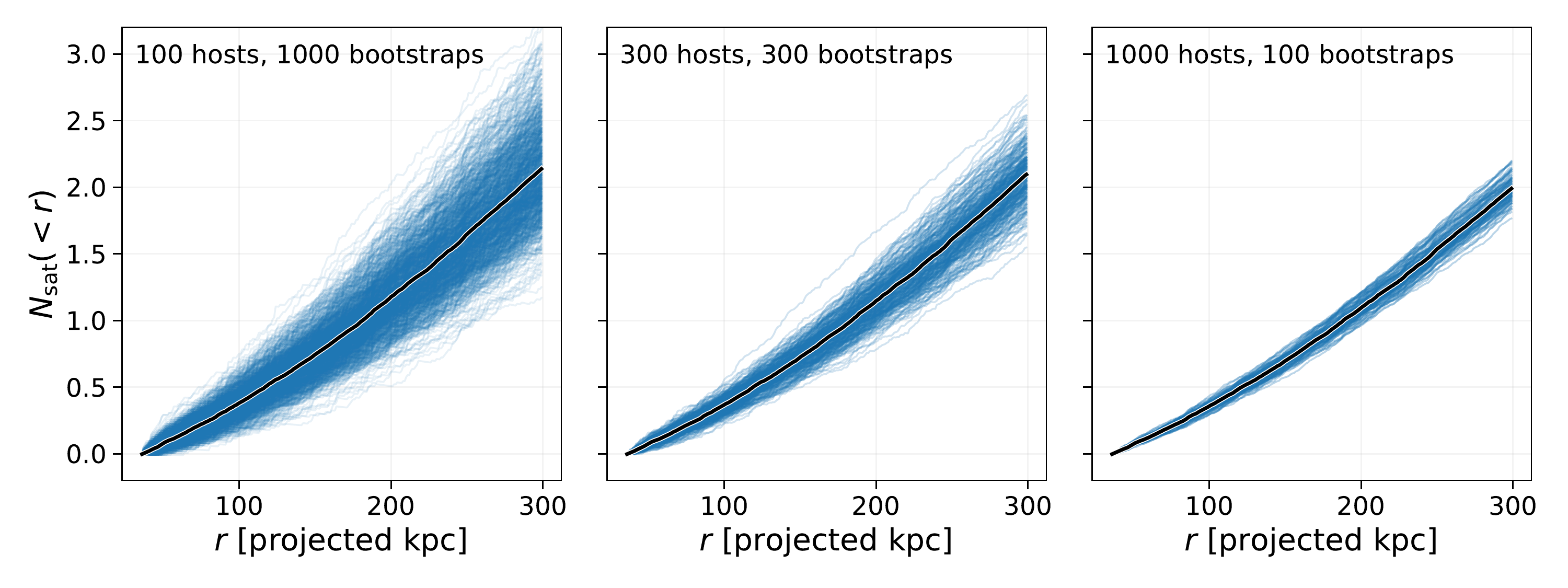}
    \caption{The radial distribution scatter varies with the number of samples drawn per bootstrap iteration. The mean cumulative number of satellites is also biased high if the number of hosts is small.}
    \label{fig:scatter}
\end{figure*}

In Figure~\ref{fig:scatter}, we report how bootstrap resampling affects the mean trend and scatter of satellite radial profiles.
In this example, we use xSAGA satellites in the projected halos of 5,334 hosts with stellar masses between $10 < \log(M_\star/M_\odot) < 11$.
We have corrected satellite number counts using the methods described in Section~\ref{sec:corrections}.

It is evident that selecting a larger number of hosts decreases the scatter of the bootstrapped profiles.
A subtle bias can be seen in the center and left-hand panels of Figure~\ref{fig:scatter}. 
The mean number of total satellites increases as a smaller number of hosts are drawn from the sample.
This bias appears because the distribution of $N_{\rm sat}(<300{~\rm projected \; kpc})$ is long-tailed; a small number of halos contain many satellites.

\bibliographystyle{aasjournal}
\bibliography{xsaga}

\end{document}